\documentclass[%
 reprint,
nofootinbib,
 amsmath,amssymb,
 aps,
]{revtex4-2}

\usepackage{graphicx}
\usepackage{dcolumn}
\usepackage{bm}
\usepackage{mathrsfs}
\usepackage{amsthm}
\usepackage{bigints}
\usepackage{xcolor}
\usepackage{hyperref}

\begin{document}

\preprint{APS/123-ART}

\title{Effective Goldstone dynamics on cosmological space-times}

\author{Marc Schneider}%
 \email{marc.schneider@ehu.eus}
\affiliation{
Department of Physics and EHU Quantum Center, University of the Basque Country UPV/EHU,\\
Barrio Sarriena s/n, Leioa 48940, Spain}
	%




\date{\today}

\begin{abstract}
We derive the Lehmann-Symanzik-Zimmermann
reduction formalism for a massive spin-2 particle on Minkowski spacetime and 
extend the formalism to cosmological spacetimes. The reduction formalism allows for a versatile proof that the Goldstone boson equivalence theorem holds in Friedmann-Lema\^itre-Robertson-Walker space-times. For the de Sitter and the radiation filled universe, we investigate the Goldstone dynamics and perform an analysis of the range of validity provided by the effective kinetic operator.
\end{abstract}

\pacs{Valid PACS appear here}
\maketitle


\section{\label{sec:level1}Introduction}

The Goldstone boson equivalence theorem (GBET) links high-energy physics 
to the longitudinal degrees of freedom. Its foundation can be linked to the applicability of the
Lehmann-Szymanzik-Zimmermann (LSZ) reduction formalism \cite{lehm57}, an operational
method to calculate $S$-matrix elements by cutting external states.
Intuitively, the GBET can be summarized as follows:
whenever the energy scale of a scattering process $E$ is much 
larger than the mass of the external state $m$, the corresponding 
$S$-matrix elements involving external, 
longitudinally polarized gauge bosons are well approximated (up to $\mathcal{O}(m/E)$-corrections)
by those containing their corresponding Goldstone scalars.  In other
words, if we have a process with a longitudinally polarized (tensor)
field, we can replace it by its corresponding scalar Goldstone partner. With this substitution, we reproduce 
the correct values for scattering amplitudes to high accuracy in the high-energy regime.
The astonishing fact is that Goldstone bosons themselves 
do not represent physical degrees of freedom; they can rather be arranged to 
disappear from any theory by gauge transformations or intrinsic cancellations.
However, despite not appearing in the external spectrum,
the Goldstone bosons still influence physical observables.

An intuitive understanding of the mechanism in the GBET can be inferred from
considering a massive spin-1 particle in the rest frame. There, its three
polarizations are approximately comparable, but if the particle starts moving, the picture changes. In particular, the two transverse polarizations become negligible at high velocities,
whereas the longitudinal polarization $\varepsilon_L$ aligns with the momentum $k$ in the direction of propagation:
$\varepsilon^a\rightarrow \varepsilon^a_L=\frac{k^a}{m}+\mathcal{O}(\frac mE)$
in the limit $\mathcal{j}k\mathcal{j}\rightarrow\infty$, such that
a massive spin-1 particle behaves like its corresponding
Goldstone scalar. Effectively, the kinetic term dominates and the 
scattering is perfectly well described by the degree of freedom 
pointing into the direction of propagation. Even though it may be a purely fiducial infrared regulator, the existence of a mass as an additional scale in the theory is presumed to be a necessity for the GBET.

Not only facilitates the GBET calculations -- it undoubtedly provides one of
the most remarkable insights to modern gauge theory. For the Standard 
Model, the GBET relates the electroweak symmetry breaking mechanism with
longitudinal weak-boson scattering experiments. This is insofar significant 
as the GBET offers another criterion for probing the Higgs mechanism
that is responsible for mass creation within the Standard Model. Moreover, the GBET is
a fabulous tool to describe models
involving a heavy Higgs boson, longitudinally polarized external 
gauge bosons, and scalar particles \cite{farzinnia2015prospects,cuomo2020goldstone}.

Historically, the GBET was first formulated in 1974 by Cornwall, Levin, 
and Tiktopoulos \cite{corn74} for a tree-level process with one
external longitudinal vector boson. Later, it was generalized
by Chanowitz and Gaillard \cite{chan85}, Lee, Quigg, and Thacker \cite{lee77} 
and Gounaris, K\"ogerler, and Neufeld \cite{gou86} to an arbitrary number 
of external, longitudinal vector bosons. Their claim that the 
GBET were true for all orders in perturbation theory has been 
proven wrong by Yao and Yuan \cite{yao88} and Bagger and 
Schmidt \cite{bag90} who pointed out that at higher order in loop corrections the
renormalization condition for the unphysical Goldstone boson has to be
taken into account. 

Based on these results, there have been many
investigations on the precise formulation of the 
GBET \cite{he92,he94,he97} but so far
most research has been confined to lower spin fields like in
the Standard Model, or related chiral theories. 
However, the equivalence theorem offers advantages that go beyond 
mere computational convenience, it has been conjectured for spin-2 theories 
both on Minkowski and Friedmann-Lema\^itre-Robertson-Walker 
backgrounds \cite{ark03,berk10,berk12}.
The conjecture for validity of a spin-2 GBET basically rests upon the same
observation that leads to the hitherto equivalence theorems. 

This rather generic statement that the longitudinal term dominates at high energies
should equally apply to massive spin-2 particles since 
at high energies, the longitudinal polarization scales as $\varepsilon^L_{ab}\sim k_ak_b/m^2$ implying that scattering amplitudes are dominated by the scalar Stückelberg decomposition. In curved space-times this statement holds locally provided that the curvature is sufficiently small compared to the energy. This ensures that the curvature operators are suppressed. 

To extract the leading order contribution to 
scattering amplitudes, we focus our attention on scalar Goldstone
bosons. In this sense, the GBET represents a natural diagnostic
tool for investigating the consistency of theories involving longitudinal
degrees of freedom. In particular, the authors of \cite{ark03} found
that Goldstone bosons in Minkowski space-time become strongly coupled 
at far lower energy than na\"{\i}vely expected from unitary gauge
considerations due to obscured intrinsic cancellations.
At low energies, a proper ultraviolet complete theory of gravity requires, 
at the linearized level, a transfer 
to the Goldstone theory. On top of that, the 
Goldstone boson description helps elucidating many peculiarities 
associated with infrared deformations of gravity in an intuitive fashion:
the explicit form of the Fierz-Pauli mass term \cite{paul39} 
is uniquely determined because any deviation would cause terms with four derivatives to occur, indicating 
pathological degrees of freedom \cite{hint12}. 
The influence of the Goldstone degree of freedom develops 
the van Dam-Veltman-Zakharov discontinuity which is the disagreement
between a massless spin-2 and the massless limit 
of a massive spin-2 field coupled to the stress-energy tensor. 
In other words, a massive spin-2 field with five degrees of freedom coupled to 
the covariantly conserved stress-energy tensor yields,
in the massless limit, a massless spin-2 and a massless spin-1
field coupled to the stress-energy tensor and a spin-0 field coupled to
its trace with gravitational strength; this does not
occur for an initially massless spin-2. 
In general, relevant modifications of Einstein's gravity present 
an exciting possibility to probe the rigidity of general relativity on cosmological 
scales. Especially, proposals for massive gravity \cite{derham} or even higher spin theories
could be tested effectively from prediction
in the high-energy limit. A valid GBET simplifies calculations tremendously and provides
a window of opportunity to test the consistency of the theory by means of the longitudinal
(scalar) degree of freedom. For instance, the GBET has been shown to be valid for inflationary space-times \cite{green2024goldstone}.

The present analysis serves as an effective field theoretic statement and does not assume the existence of an ultraviolet completion of massive gravity. Throughout this article we presume that, below a characteristic cutoff scale, the theory admits a perturbative description in terms of a massive spin-2 field with well-defined helicity sectors. Furthermore, we assume the existence of asymptotic particle states such that standard scattering-theoretic methods can be employed. The derivation relies on the LSZ reduction formalism and the Stückelberg decomposition used to identify the longitudinal degrees of freedom. As a statement about an effective spin-2 field theory, the GBET remains insensitive to the details of the ultraviolet completion.

This article aims to derive the GBET on cosmological space-times in a versatile way. Section 2 will give a short introduction in
the LSZ reduction formalism and the corresponding GBET for massive spin-1 fields
on flat spacetime before we extend the formalism to spin-2. In section 3, we
present the generalization of the GBET to curved spacetimes for a spin-2
field with Fierz-Pauli mass term. Section 4 particularizes our findings to de Sitter and radiation dominated universes and we discuss the validity of the approximation. In the subsequent conclusion we present an outlook and 
emphasize applications of this theorem.
Our sign convention for the metric is mostly-plus, and we work in natural 
units, where $\hbar=c=G_N=1$.

\section{\label{sec:level11} Minkowski background}

Let us start with a brief introduction into
the Lehmann-Symanzik-Zimmermann (LSZ) reduction formalism
(cf. \cite{lehm57}, \cite{itz80} for a detailed presentation). Consider a
(bosonic) field
which transforms as an $(r,s)$-rank tensor $\digamma_{a_1...a_s}^{b_1...b_r}$ under
the diffeomorphism group. This tensor describes the system's degrees of freedom. In general, $\digamma$ may denote either a symmetric (bosonic) or an antisymmetric
(fermionic) field. In this case, the degree of freedom also carries Lorentz indices. So far, the GBET has been 
thoroughly studied for rank $(0,0)$ (scalar or simply spin-0) and rank $(0,1)$ (vector or spin-1) bosons as well as 
for fermionic degrees of freedom described via spinors \cite{itz80}.
In the following we restrict ourselves to bosonic degrees of freedom with the note that the fermionic generalization follows immediately. 

By using the two metalabels $A=\{a_1,\ldots,a_s\}$, and $B=\{b_1,\ldots,b_r\}$, which collect all the individual indices and their contractions, we will show the derivation of the LSZ formalism
schematically and then apply it to the vector boson before extending to the case of a spin-2 field described by a $(0,2)$-tensor. 
Our starting point is a scattering process with one 
ingoing rank-$(B,A)$ quantum field $\digamma_{\rm in}(k,\varepsilon)$ described by the
creation operator $a^{\dagger}_{\rm in}(k,\varepsilon)$ 
acting on the in-state $|\psi_{\rm in}\rangle$.
We expand the field operator into positive frequency modes and their hermitean conjugates
\begin{equation}
\digamma^{B}_A(x)=\underset{n}{\sum}\left(a_{n}(k)(f^B_{A,n}
(k;x))^*+a_{n}^{\dagger}(k)f^B_{A,n}(k;x)\right)
\end{equation}
where the polarization tensor $\varepsilon_{A,n}^B(k)$ has been absorbed into the
basis $f_{A,n}^B(k,x)=\varepsilon_{A,n}^B(k)f(k;x)$
which is given by plane waves $f(k;x)=e^{ikx}$ in flat space-time. Note, $x$ denotes the space-time location such that the exponent has to be understood as $kx:=k_ax^a$. We can now 
perform a shift in the momentum to change to a four-dimensional integral and write \cite{itz80}
\begin{equation}
\begin{aligned}
\mathcal{j}\mathrm{\psi_{\rm in}}\rangle = & \;i\int\mathrm{d}^{4}x\,\partial^{c}
\left(\digamma^B_A\overset{\leftrightarrow}{\partial}_{c}(f_B^A(k;x))^*\right)\mathcal{j}\mathrm{\psi_{\rm in}}-k\rangle\\
 & -i\underset{t\rightarrow+\infty}{\lim}\int\mathrm{d}x^4n^c\, \digamma^B_A\overset{\leftrightarrow}{\partial}_{c}(f_B^{A}(k;x))^*\mathcal{j}\mathrm{\psi_{\rm in}}-k\rangle\label{a}
\end{aligned}
\end{equation}
where $n^a$ is the normal vector while the left and right anti-symmetric derivative operator is defined as
\begin{equation}
    f(x)\overset{\leftrightarrow}{\partial}_ag(x)=f(x)(\partial_ag(x))-(\partial_af(x))g(x)
\end{equation}
Note, we have used 
asymptotic freedom for $\digamma_A^B$, that is, for $t\to\mp\infty$ the fields become effectively free, i.e.
\begin{equation}
    \digamma_A^B\to Z\digamma_A^{B,{\rm in/out}}
\end{equation}
where the normalization $Z$ of the field $0<Z\le1$ for which equality occurs in the free field case\footnote{The constant $Z$ is subject to a normalization procedure with respect to the equal time commutator. It takes into account that the field content $\digamma_A^B(x)|0\rangle$ is not fully covered by $\langle 1|\digamma_A^B(x)|0\rangle$ whilst it would be by using $\digamma_A^{B,{\rm in}}(x)$ instead \cite{itz80}.}. Physically, this conditions relates quantum fields to free one-particle states at the asymptotic boundaries that are elements of the free-field vacuum's Fock space.

Since, essentially, the LSZ reduction formalism amputates the external legs in scattering amplitudes, it, in fact, connects the measurements from scattering 
amplitudes, which describes how $N$ initial particles scatter into $M$ final particles, 
with the dynamical content of the theory, i.e. the degrees of freedom in the propagator.
For the amplitude transferring an initial into a final state, we get
\begin{equation}
\begin{aligned}
\mathcal{M}=\langle\psi_{\rm out}\mathcal{j}\psi_{\rm in}\rangle = \;& \mathrm{disc.}+\nonumber 
 i\int\mathrm{d}^{4}x\, (f_B^A(k_1;x))^*\,\mathfrak{D}^{BC}_{AD}(x_{1}) \\
  & \times\mathcal{h}\psi_{\rm out}\mathcal{j}\mathcal{T}[\digamma_C^D(k_{1})]\mathcal{j}\psi_{\rm in}\mathcal{i}\label{b}
\end{aligned}
\end{equation}
with $\mathfrak{D}_{ABCD}$ being the differential operator that constitutes the equation of motion for $\digamma^{AB}$
and $\mathcal{T}$ the time ordering which is included 
in case there exist more than one field. Notice, we used once again that  asymptotically $\digamma_A^B$ represents a free one-particle state. The time-ordering then ensures causality if more than one operator is involved.
Here, the disconnected terms 
(denoted by disc.) describe processes where not all participating
particles scatter\footnote{Those terms are proportional to Dirac $\delta$-distributions, explicitly
for one non-interacting particle with initial momentum $q_1$ we get
\begin{equation}
\mbox{disc.}\!=\!\sum_{k=1}^M(2\pi)^{3}p_k^0\delta^{(3)}(q_1-p_k)\langle p_1,...,p_{k-1},
p_{k+1},...|q_2,...\rangle
\end{equation}
where the summation extends over all possible out-states. Here, $p$ and $q$ denote the momenta.}.
Since the addition of a spectator particle should not influence the GBET, these terms are not essential and will be neglected for further considerations.
 
It is important to mention that the LSZ formalism
requires the presence of a (fiducial) mass; this is because mass terms break gauge invariance. Thus, a formerly spurious degree of freedom becomes observable; in case of a vector
boson a longitudinally polarized mode will enter the physical spectrum. 
By a standard argument in differential geometry, $\varepsilon_A^B$ can be split into a tensor product of $A$ polarization vectors and $B$ co-vectors, such that $\varepsilon=(\bigotimes_A\varepsilon) \otimes(\bigotimes_B\varepsilon^*)$. Recall the Ward-Takahashi identity for amplitudes $\mathcal{M}=\varepsilon^A_B\mathcal{M}_A^B$, scaling
relations for high energies unveil
that the dominant contribution is described by the polarization which is mostly
aligned with the propagation direction. This allows us to replace the polarization tensor as follows
$\varepsilon_A\rightarrow \frac{1}{m^r}\prod_{i=1}^r k_{a_i}$. As a result, we get for the 
reduced scattering amplitude (for the sake of simplicity only for one momentum)
\begin{equation}
k_{b}\mathcal{M}^{b} = i\!\int\! {\rm d}^4x\;e^{ikx} \partial_{b}\mathfrak{D}^{B(b)}_A(x_{1})\langle\psi_{\rm out}\mathcal{j}\mathcal{T}\{\digamma_{B}^A(x_{1})\}\mathcal{j}\psi_{\rm in}\rangle \label{eq:1lszred}
\end{equation}
where we have pulled $k_b$ under the Fourier integral such that $k_a\to i\partial_a$. We will specify the explicit form of $\mathfrak{D}$ in $\mathcal{M}^a$ in the next sections which cover the specific examples of the tensor fields. 
This rest matrix element covers the amputated scattering diagrams and will therefore be
central in the derivation of the GBET. 
  
\subsection{The GBET for Proca theory}\label{s1}  

To illustrate the methodology, we begin with a well-known theory to study the effect as well as the apparent physical 
relevance of the Goldstone
boson. Consider a massive vector (or a $(0,1)$-tensor) field $\tilde{A}$ that is described by Proca theory.
This theory offers the right amount of complexity while being essential in the Standard Model for the description of the massive gauge bosons,
i.e. $W^\pm$ and $Z^0$, of the weak interaction sector.
The Lagrange density is given by
\begin{equation}
\tilde{\mathcal{L}}_{P} = -\frac{1}{4} F^{ab}(\tilde{A}) F_{ab}(\tilde{A})
+ \frac{1}{2} m^2 \tilde{A}_a\tilde{A}^a
 + J^a\tilde{A}_a.\label{2.34}
\end{equation}
with rank $(0,2)$ field strength tensor $F={\rm d}\tilde{A}$ and external current $J$.
The vector field $\tilde{A}$ propagates three physical degrees of freedom: one 
longitudinal scalar degree of freedom and two transversal
vector degrees of freedom. 
The kinetic operator for Proca theory expressed in the $R_\xi$-gauge \cite{itz80}
\begin{equation}
\mathcal{D}_{ab}=(\Box-m^2)\eta_{ab}+\left(\frac1\xi-1\right)
\partial_a\partial_b\label{ProcK}
\end{equation}
where $\eta$ is the Minkowski metric and $\Box:=\eta^{ab}\partial_a\partial_b$ the d'Alembert operator. Let us emphasize that for our methodology it will be essential to work in this gauge. 

If we compare Proca theory to its massless counterpart, Maxwell theory,
we see that the mass term breaks the gauge 
freedom explicitly such that an additional degree of freedom arises. As pointed out in \cite{ark03}, 
gauge freedom corresponds to the existence of spurious degrees of freedom $s$
in the theory. Those are in principle part of the theory but can not contribute to 
physical observables since $\|s\|=0$ and hence for any operator $\mathcal{O}$ follows
$\langle s|\mathcal{O}|s\rangle=0$ \cite{gsw1}. Although they are present in the 
theory, we might not see their influence as long as the corresponding gauge symmetry is active. Observables will stay unaffected from those degrees of freedom such that before making any prediction, we need to fix a gauge.
In other words, whenever we break a gauge freedom, we activate a spurious degree of freedom. Exactly 
this happens through addition of a (fiducial) mass term in a massless theory. To turn it around, we 
can also artificially amend the gauge freedom by introducing a St\"uckelberg field $\phi$ through the transformation:
\begin{equation}
\tilde{A}_a\longrightarrow A_a+\frac{1}{m}\partial_a\phi\label{2.36}.
\end{equation}
As we see immediately, the existence of a mass term in $\mathcal{L}$ is essential for this formalism. The mass $m$ is introduced because $\phi$ is canonically normalized.
The St\"uckelberg field $\phi$ plays the role of a Goldstone boson and now 
\eqref{2.34} exhibits that the artificially created symmetry grants us some additional gauge freedom.
In fact, we have included a helicity-1 shift symmetry $\delta A_a=\partial_a \chi$ which can be 
restored by a shift in the St\"uckelberg field $\delta \phi=-m\chi$.
Using this, we implement the following $R_\xi$-gauge-fixing condition by adding\footnote{The gauge fixing conditions and the symmetries are closely related to the BRST transformation \cite{becch76} which is explicitly discussed for the case of the standard model in \cite{hor97}.}
\begin{equation}
\mathcal{L}_{\rm gf}=-\frac{1}{2\xi}\left(\partial_a A^a+\xi m\phi\right)^{2}\label{eq:generalgf}
\end{equation}
to $\mathcal{L}_P$ with arbitrary gauge parameter $\xi$ implying
$\langle\psi\mathcal{j}\partial_a A^a+\xi m \phi\mathcal{j}\psi\rangle=0$, that is, for physical states, the Stückelberg Lagrange density equals the Proca Lagrange density. Although in principle \eqref{eq:generalgf} may lead to pathologies, Delbourgo \emph{et al.} \cite{delbourgo1988massive} have shown that the theory does not feature ghost states that couple to physical states.
 
Before proceeding, we need to check whether this gauge condition can be consistently imposed, that is, if it can be reached by an appropriate gauge transformation. If there were no additional St\"uckelberg field $\phi$, we could choose the
Lorenz gauge $\partial_a\tilde{A}^a=0$. However, using \eqref{2.36} we find the condition
\begin{equation}\label{gaugeM}
\partial_a A^a+\frac{\Box\chi}{m}=0\quad\Rightarrow \quad\Box\chi=-m\partial_a A^a.
\end{equation}
The gauge is compatible with the theory whenever there exists a solution for $\chi$ to 
\eqref{gaugeM} under consideration of the inhomogenous Gupta-Bleuler condition. The gauge condition implies the constraint 
\begin{equation}\label{eq:GBEB}
    \partial_a A^a=-m\phi
\end{equation}
which removes the time-like component but leaves the longitudinal mode dynamical\footnote{In classical electrodynamics this would correspond to an
inhomogenous Lorenz gauge which eliminates only the temporal entry in $A_a$ but not
the one parallel to the direction of propagation.}.
Now $\chi$ has to fulfill just the Klein-Gordon equation which with the gauge condition \eqref{eq:GBEB} becomes
\begin{equation}
    \Box\chi=\xi m^2\phi.
\end{equation}
Its solution
is known to be given by means of the fundamental solution $G(x,x')$ as
\begin{equation}
\chi(x)=-m^2\xi\int_{\mathcal{M}}{\rm d}^4x'G(x,x')\phi(x'),
\end{equation}
where $G(x,x')$ denotes a Green’s function of the d’Alembert operator (e.g. the retarded Green’s function), which exists on globally hyperbolic space-times.
Hence, equation \eqref{eq:GBEB} defines a valid gauge fixing because every field configuration can be mapped into it.

Let us have a look at the polarization vector in this theory: $\varepsilon_a$ is normalized such that $\|\varepsilon\|=-1$. A canonical basis can be found by choosing the 
propagation direction such that $k_a$ points in the $z$-direction, with $E^2-k_z^2=m^2$.
Imposing the Gupta-Bleuler quantization condition $\varepsilon_a k^a=0$ has to be fulfilled. 
The two obvious choices for the transversal polarizations are found to be $\varepsilon_a^{T_1} 
=(0,1,0,0)$ and $\varepsilon_a^{T_2} =(0,0,1,0)$ while the third 
polarization is $\varepsilon^L_a=(\frac{k_z}{m},0,0,\frac Em)$. Note, $\varepsilon_a^L$ is the
longitudinal polarization. Its energy dependence is a direct consequence of the little group and constitutes the correspondence between $\varepsilon^L_a$ and 
$\frac{k_a}{m}$.

In order to extract the kinetic operator of  $A_a$  from 
$\mathcal{L}_{P}+\mathcal{L}_{\rm gf}$,
employ the LSZ reduction formula as well as the condition that for high-energies
$\varepsilon\to \frac km+\mathcal{O}(\frac mE)$ which projects out the longitudinal degree of 
freedom $A^L$. 
However, we follow a different, more straightforward
We use a very direct approach by applying the derivative to the kinetic operator. Then, we are able to derive the GBET for one spin-1 particle in the in-state \cite{itz80}: 
\begin{equation}
\begin{aligned}
\varepsilon^{a}&\mathcal{M}_{a}(\psi_{\rm in}+ A_{a}(k, \varepsilon)\rightarrow\psi_{\rm out} )\\
&\!\overset{\eqref{ProcK}}{=}\frac{i}{m}\int\!\mbox{d}^4x \;e^{-ikx}\partial^a\!\left[(\partial^c\partial_c-m^2)\eta_{ab}+\left(\frac1\xi-1\right)\partial_a\partial_b
\right]\\
&\hspace{1.3cm}\times\langle \psi_{\rm out}|A_b(x)|\psi_{\rm in}\rangle\\
&=-\frac{i}{m}\int\!\mbox{d}^4x\;e^{-ikx}\left(\frac{\Box}{\xi}-m^2\right)
\langle \psi_{\rm out}|\partial_aA^a(x)|\psi_{\rm in}\rangle\\
&\!\overset{\eqref{eq:generalgf}}{=}-i\int\!\!\mbox{d}^4x\,e^{-ikx}(\Box-m^2)\langle \psi_{\rm out}|\phi(x)|\psi_{\rm in}\rangle\\
&=-i \mathcal{M}(\psi_{\rm in}+\phi(x)\rightarrow\psi_{\rm out})
\left[1+\mathcal{O}\left(\frac{m}{E}\right)\right]
\label{2.55}
\end{aligned}
\end{equation}
where we used \eqref{eq:GBEB} in the last step to replace $\partial_aA^a(x)$ by $\xi m\phi(x)$. After canonically normalizing $\phi$ by $\phi/m$, we find the Klein-Gordon equation for the Goldstone scalar and thus the GBET for Proca. Let us emphasize that in our methodology it was essential to work in the $R_\xi$-gauge. This is because we do not intend to enforce certain conditions a priori. 
The proof of the GBET in Proca theory used Schwarz's theorem, that is, partial derivatives commute, as such the differential operator \eqref{ProcK} commutes with $\partial_a$. Application of
$\mathcal{D}^{ab}$ to $A_a$ will not give zero because $A_a$ is not an asymptotic
state instead it solves $\mathcal D^{ab} A_a=J^b$ for $J^b\neq0$. Finally, the differential operator 
is transformed to the Klein-Gordon operator for the scalar field, which look similar here. However, we
will see examples, where this is not the case, especially in curved space-times. 

The term marking
the low-energy corrections comes from the transversal contributions which are ignored
in the replacement of the polarization tensor by the longitudinal wave-vector.
A generalization to arbitrary numbers of spin-1 particles in 
the in- and out-states can be accomplished in a similar way \cite{chan85}. Intuitively,
from \eqref{2.55} follows that in the high energy limit the amplitude for a $(0,1)$-field
is equivalent those of a $(0,0)$-field up to $\mathcal{O}(\frac mE)$.

If we had instead made use of a longitudinal projector 
$\mathcal{P}_L$ applied onto $\mathcal{D}$, this would have yielded an equation of motion that is
solved directly by $A^L$. The longitudinal degree of freedom by itself is a scalar and thus obeys the Klein-Gordon equation. In other words, \eqref{ProcK} applied to $A^L_a:=(\mathcal{P}_LA)_a$ would automatically reduce such that $(\mathcal{P}_L\mathcal{D})_{ab}=(\Box-m^2)\eta_{ab}$ from which the GBET follows immediately: after applying the derivative, we use the Gupta-Bleuler gauge condition and replace $\partial_aA^a$ by $\phi$ which agrees with the result before.

\subsection{\label{sec:level22}The GBET for massive spin-2 fields}

To extend the GBET to spin-2 bosons, we will essentially
follow the steps outlined in the previous section.
Our starting point is the theory of a free $(0,2)$-tensor field
$h_{ab}$ which acquires a mass from the Fierz-Pauli
term \cite{ark03}
\begin{equation}
\begin{aligned}
\mathcal{L}=&-\frac{1}{2}\partial_a\tilde{h}_{bc}\partial^{a}\tilde{h}^{bc}+\partial_{b}\tilde{h}_{ca}\partial^{c}\tilde{h}^{ba}-\partial_{b}\tilde{h}^{bc}\partial_{c}\tilde{h}\\
&+\frac{1}{2}\partial_{a}\tilde{h}\partial^{a}\tilde{h}-\frac{1}{2}m^2(\tilde{h}_{ab}\tilde{h}^{ab}-\tilde{h}^2)\label{2.75}.
\end{aligned}
\end{equation}
with $h=tr_\eta(h_{ab})$ and $m$ the mass of $h_{ab}$.
A massive spin-2 field propagates five degrees of freedom and one ghost degree of freedom, the Boulware-Deser ghost \cite{bd72} which disappears at the linearized level but reappears in the non-linear theory. This instability comes from the loss of the Hamiltonian 
constraint, however, it has been shown that non-linear theories without Boulware-Deser instability can be constructed by imposing a second
constraint \cite{hass12}. 

The kinetic operator $h_{ab}\mathscr{D}^{abcd}h_{cd}$ can be retrieved after an integration by parts of \eqref{2.75}, thus we find the explicit form of the Lichnerowicz operator ${\mathscr{D}^{ab}}_{cd}$ to be
\begin{equation}
\begin{aligned}
&{\mathscr{D}^{ab}}_{cd}=\frac{1}{2}\left(\Box-m^2\right)\left(\frac{1}{2}{\eta^a}_c{\eta^{b}}_d+\frac{1}{2}{\eta^{a}}_d{\eta^{b}}_c-\eta^{ab}\eta_{cd}\right)\\
&+\eta^{ab}\partial_{c}\partial_{d}+\eta_{cd}\partial^{a}\partial^{b}\\
&-\frac{1}{2}\left({\eta^b}_d\partial^{a}\partial_{c}+{\eta^b}_c\partial_d\partial^{a}+{\eta^a}_d\partial^{b}\partial_c+{\eta^a}_c\partial^{b}\partial_d\right)\label{2.81}.
\end{aligned}
\end{equation}
Similar to Proca theory, we enhance our symmetry using
the St\"uckelberg formalism which will then culminate into the Goldstone vectors and scalars
for the massive $(0,2)$-tensor field. For the sake of clarity, 
we will perform the St\"uckelberg decomposition step by step
starting with the introduction of Goldstone vectors $\tilde B_a$
\begin{equation}
\tilde{h}_{ab}\longrightarrow h_{ab}+\frac{1}{2m}\partial_{a}\tilde B_{b}+\frac{1}{2m}\partial_{b}\tilde B_{a},\label{2.76}
\end{equation}
The gauge symmetry is hence given by the transformations $\delta h_{ab}=\frac{1}{2}(\partial_a \zeta_b
+\partial_b\zeta_a)$
and the shift in the Goldstone degree of freedom $\delta \tilde B_a=-m\zeta_a$.
The redundancy \eqref{2.76} will produce mixed
 terms in $\mathcal{L}$ which we wish to eliminate by adding a suitable $R_\xi$-gauge
\begin{equation}
\mathcal{L}_{\rm gf}=\frac{1}{2\xi}\left(\partial^{a}h_{ab}-\partial_{b}h-\xi m\tilde B_{b}\right)^2\label{2.78}
\end{equation}
where the square shall be understood as a contraction in the free indices. As such, $\mathcal{L}$ is diagonal in the fields $h_{ab}$ and $\tilde B_{a}$.
The above gauge condition could be viewed as an inhomogeneous de Donder-gauge with the 
Goldstone boson as sourcing term; this is similar to the Proca case where the scalar field
was sourcing the inhomogeneous Lorenz gauge. Since the massless part of \eqref{2.75} admits gauge invariance, only the Fierz-Pauli term changes and we get an 
additional term for the Stückelberg fields.
Note, $\mathscr{D}_{abcd}$ contains a longitudinal part which is directly proportional to the
Klein-Gordon equation, the other terms describe transversal degrees of freedom. On this level, it becomes evident, that if the longitudinal degree of freedom
dominates, then, we will be left with a scalar field theory. By reducing the theory further,
we will eventually arrive at a theory that incorporates a scalar degree of freedom aka the 
longitudinal Goldstone mode.

The first step to show the GBET employs the LSZ formalism to reduce a $(0,2)$-tensor field in the outgoing state. The rest matrix element (cf. section \ref{s1}) reads
\begin{equation}
\begin{aligned}
\mathcal{M}_{ab}(\psi_{\rm in}\rightarrow \psi_{\rm out} + h_{ab})=&\;C_h\int{\rm d}^4x\;e^{ikx}\\
&\times
{\mathscr{D}_{ab}}^{cd}
\mathcal{h}\psi_{\rm out}\mathcal{j}h_{cd}\mathcal{j}\psi_{\rm in}\mathcal{i}\label{2.83},
\end{aligned}
\end{equation}
where $C_h$ collects all constants because they are irrelevant for the 
GBET's mechanism. Hence, we will absorb all constants that appear in the course of the proof into $C_h$ without
further notice.

Following the steps in the previous section,
we use the Ward-Takahashi identity and split the matrix element into polarization vectors and 
rest-matrix element 
$\mathcal{M}\rightarrow\mathcal{M}_{ab}\varepsilon^{ab}$. After splitting up the polarization tensor into two polarization vectors, we
perform the high-energy replacement of one $\varepsilon^a$, such that,  $\mathcal{M}_{ab}$ is multiplied with a momentum $k^{a}/m$. Then, we 
recall the general Ward identity, and find the equation for the longitudinal mode of the Goldstone vector
\begin{equation}
\begin{aligned}
-i\frac{k^{a}}{m}&\mathcal{M}_{ab}(\psi_{\rm in}\rightarrow \psi_{\rm out} + h_{ab})\\
=&\frac{C_h}{m}\int {\rm d}^4x\;e^{ikx} \partial^{a}{\mathscr{D}_{ab}}^{cd}
\mathcal{h}\psi_{\rm out}\mathcal{j}h_{cd}\mathcal{j}\psi_{\rm in}\mathcal{i}\\
=&-\frac{C_h}{m}\int {\rm d}^4x\;e^{ikx}\left(\frac{\Box}{\xi}-m^2\right)
\mathcal{h}\psi_{\rm out}\mathcal{j}m\xi\tilde B_{b}
\mathcal{j}\psi_{\rm in}\mathcal{i}
\end{aligned}
\end{equation}
for the first step (cf. Appendix \ref{appb} for details). 
Here, we used the gauge condition to replace
$\partial^{a}h_{ab}-\partial_{b}h=m\tilde B_{b}$
derived from (\ref{2.78}). This result holds of course in leading order of the high-energy limit where the polarization tensor can be approximated by
a product of polarization vectors which high-energy behavior is 
dominated by the momentum
($\varepsilon^{a}_{L}\approx\frac{k^{a}}{m}+\mathcal{O}(\frac Em)$). Therefore
we get for the first reduction
\begin{equation}
i\frac{k^{a}}{m}\mathcal{M}_{ab}(\psi_{\rm in}\rightarrow \psi_{\rm out}+h_{ab})
=\mathcal{M}_{b}(\psi_{\rm in}\rightarrow \psi_{\rm out}+\tilde B_{b})\label{2.84}
\end{equation}
We proceed as just before, however, with the new starting point given by the Lagrange density for the vector field
\begin{equation}
\mathcal{L}=-\frac18F^{ab}F_{ab}+\frac{\xi}{2}m^2\tilde B^2
\end{equation}
Our next step is to include a new degree of freedom $\phi$
via the St\"uckelberg transformation
\begin{equation}
\tilde{B}_{a}\longrightarrow B_{a}+\frac{\partial_{a}\phi}{m}\label{2.86}.
\end{equation}
The scalar $\phi$ will play the role of the Goldstone scalar and again induces mixing terms 
in the Lagrange density which are removed by a second gauge-fixing similar to the inhomogenous
Gupta-Bleuler condition in the $R_\xi$-gauge
\begin{equation}
\mathcal{L}_{\rm gf2}=\frac{1}{2\lambda}(\partial_{a}B^{a}-\lambda m\phi)^2.\label{2.85}
\end{equation}
Addition of \eqref{2.85} under consideration of \eqref{2.86}
yields the desired diagonal form
\begin{equation}
\mathcal{L}+\mathcal{L}_{\rm gf2}=B_{a}\mathcal{D}^{ab}B_{b}+\phi D\phi\label{2.87},
\end{equation}
where ${\mathcal{D}_{ab}}$ and $K$ in
 \eqref{2.87} are the kinetic operators with respect to the fields 
 $B^{a}$ and $\phi$ (with $\lambda\equiv1$)
\begin{eqnarray}
&\mathcal{D}_{ab}=\frac{1}{4}\Box\eta_{ab}-\frac{1}{4}\partial_a\partial_b+\frac{1}{2}m^2\eta_{ab}\\
&D=-\frac{1}{2}(\Box+m^2).
\end{eqnarray}
To reduce the scattering matrix element, we exploit again the Ward-Takahashi identity and multiply the 
scattering matrix element with a momentum $k^{a}$. 
Recalling the formula, we have already evaluated for spin-1 (cf. \ref{s1}), we get
\begin{equation}
\begin{aligned}
-i\frac{k^{a}}{m}&\mathcal{M}_{a}(\psi_{\rm in}\rightarrow \psi_{\rm out}+ B_{a})\\
&=C_B\int{\rm d}^4x\;e^{ikx}\frac{-i}{m} \partial_{a}\mathcal{D}^{ab}
\mathcal{h}\psi_{\rm out}\mathcal{j}B_{b}\mathcal{j}\psi_{\rm in}\mathcal{i}\\
&=C_B\int {\rm d}^4x\;e^{ikx}\frac{i}{m}\mathcal{h}\psi_{\rm out}\mathcal{j}m\Box\phi-\lambda m^3\phi
\mathcal{j}\psi_{\rm in}\mathcal{i}\\
&=
\mathcal{M}(\psi_{\rm in}\rightarrow \psi_{\rm out}+\phi).
\end{aligned}
\end{equation}
In the last step, we inserted the high-energy approximation $\frac{k^{a}}{m}\approx \varepsilon^{a}_{L}$. Putting everything together we have proven the GBET for massive spin-2 
fields. The relation between the longitudinal degree of freedom in $h_{ab}$ and the 
Goldstone scalar can be written in terms of the rest-matrix $\mathcal{M}_{ab}$ for 
$h_{ab}$ and the amplitude $\mathcal{M}$ for $\phi$ as
\begin{equation}
\begin{aligned}
\varepsilon^{a}_{L}&\varepsilon^{b}_{L}\mathcal{M}_{ab}
(\psi_{\rm in}\rightarrow \psi_{\rm out} + h_{ab})
\simeq\varepsilon^{a}_{L}\mathcal{M}_{a}(\psi_{\rm in}\rightarrow \psi_{\rm out}+ B_{a})\\
&\simeq\mathcal{M}(\psi_{\rm in}\rightarrow \psi_{\rm out}+\phi)\left[1+\mathcal{O}\left(\frac{m}{E}\right)\right].
\end{aligned}
\end{equation}
which successively confirms the GBET for spin-2 tensor fields in the 
high-energy limit on a flat background.

Although direct experimental measurements of graviton scattering amplitudes are still inaccessible, the $S$-matrix facilitates the probing of the high-energy structure in massive spin-2 theories. In particular, the GBET shows that the helicity-0 sector is the dominant degree of freedom at high energies which is also responsible for many distinctive features of massive gravity, such as strong-coupling phenomena, decoupling limits, and potential instabilities.

\section{\label{sec:level111}Curved backgrounds}

Compared to Minkowski spacetime, curved spacetimes unveil a vast variety of subtleties pertinent to quantum field theory. To extend our results, we need a brief investigation of these peculiarities which mostly root in the inapplicability of the Stone-von Neumann theorem. 
We recall that fields on curved space-times are not representations
of the Poincar\'e group \cite{dew75}
that is to say, there are infinitely many unitarily nonequivalent vacuum states; none of which
needs to be preferred. In de Sitter spacetime, the Bunch-Davies vacuum represents such a preferred vacuum, because it fulfills the Hadamard condition \cite{dap09,dmor09}, i.e. the light-cone
approaches the Minkowski light-cone, while all isometries of the 
background are respected \cite{ag15}.

There exists another difficulty that arises from the lack of Poincar\'e symmetry, the Hilbert spaces at infinite past and infinite future are distinct, i.e. 
$\mathscr{H}_{\rm in}\neq\mathscr{H}_{\rm out}$.
Therefore, the states
in the initial Hilbert space $\mathscr{H}_{\rm in}$
have to be related with the final Hilbert
space $\mathscr{H}_{\rm out}$ via a Bogolubov transformation. In other words, even for free fields, the evolution may involve a scattering operator due to the nonequivalence of vacua \cite{ash75}. Intuitively, curved spacetime represents an external quantity 
which is coupled to the specific quantum field theory. This 
motivates the presence of a scattering operator $\mathcal{S}:\mathscr{H}\to
\mathscr{H'}$
even in persistency amplitudes in order to
connect both Hilbert spaces, i.e. $\langle\psi_{\rm out}|\psi_{\rm in}\rangle=
\langle\psi_{\rm in}\mathcal{S}|\psi_{\rm in}\rangle$. The geometric information
is encoded in the Bogoliubov transformation and in principle also in the vacuum \cite{bir80}.
Another peculiarity comes with the definition of the Fourier transform. Flat space-time
descriptions provide plane waves as harmonics which in turn serve as complete basis
for a Fourier expansion. In curved space-time, this procedure can not be applied in general. For
space-times with certain symmetries, like Bianchi-I space-times we could use plane waves but e.g. for spherically symmetric space-times we would need spherical hamonics as basis.

Our starting point is the Lagrangian density for a massive
spin-2 particle on an arbitrarily curved background
\begin{equation}
\mathcal{L} = \mathcal{L}_{EH}-\frac{1}{2}m^{2}(\tilde{h}_{ab}\tilde{h}^{ab}
-\tilde{h}^{2})-\frac{1}{2}\tilde{h}_{ab}T^{ab}
\end{equation}
where $\mathcal{L}_{EH}$ denotes the Einstein-Hilbert part of the Lagrangian which is the 
kinetic part for $h_{ab}$. Massive gravity theories have been consistently formulated on curved space-times by 
de Rham et. al \cite{derham} who showed that the theory can 
be resummed and is ghost-free at all orders of the decoupling limit. 

At this point, we take the position of studying a spin-2 field regardless if it describes gravity. 
Following the steps in \ref{sec:level22} we introduce new degrees
of freedom through the St\"uckelberg technique,
\begin{equation}
\tilde{h}_{ab} \rightarrow h_{ab}+\frac{\nabla_{a}A_{b}}{2m}+\frac{\nabla_{b}A_{a}}{2m}\label{eq:st=0000FCck1}
\end{equation}
with $A_{a}$ being a St\"uckelberg vector and $m$
the mass. Note, the derivative is replaced by a covariant 
derivative $\nabla_av_b=\partial_av_b+\Gamma_{ab}^cv_c$ with $\Gamma_{ab}^c=\frac{1}{2}g^{cd}\partial_b g_{ad}$
the Christoffel symbol. The Lagrange density $\mathcal{L}_{EH}$ is not
invariant under the St\"uckelberg transformation,
and thus yields additional terms proportional to the vector field $A_{a}$ 
and the Ricci curvature tensor $R_{ab}$. However, since the whole
theory is gauge invariant, these terms cancel with terms
coming from the coupling to the source $\tilde{h}_{ab}T^{ab}$.

Introducing St\"uckelberg fields provides additional
symmetries \cite{ark03} which provide a gauge freedom we use to bring the Lagrangian
into diagonal form. We choose as gauge-fixing condition the following  
 \begin{equation}
\mathcal{L}_{\rm gf}=\frac{1}{2\lambda}\left(\nabla^{a}h_{ab}-\nabla_{b}h-\lambda mA_{b}\right)^{2}.\label{eq:gf}
\end{equation}
By adding this gauge-fixing to our original Lagrangian 
we will be able to extract the kinetic operator 
$\left({\mathscr{D}^{ab}}_{cd}\right)_{x}$. 
The explicit form of 
$\left({\mathscr{D}^{ab}}_{cd}\right)_{x}$ can be found to be (cf. Appendix \ref{appa} for the explicit form of the Lagrange density)
\begin{equation}
\begin{aligned}
&\mathscr{D}_{abcd} = -\frac{1}{2}m^{2}\left(\frac{1}{2}g_{ac}g_{bd}+\frac{1}{2}g_{bc}g_{ad}-g_{ab}g_{cd}\right)\\
&-\frac{1}{2\lambda}g_{ab}g_{cd}\Box
+\frac{1}{16\lambda}R_{(ac}g_{bd)}
 -\frac{1}{16\lambda}g_{(ad}\nabla_{c}\nabla_{b)}\\
 & +\frac{1}{8\lambda}\left(R_{dabc}+R_{dbac}\right)+\frac{1}{4\lambda}g_{(ab}\nabla_{c}\nabla_{d)}.\label{eq:44-1}
\end{aligned}
\end{equation}
Here we used the usual convention for the d'Alembert operator $\Box=\nabla_a\nabla^a$
where the contraction is performed by the background metric $g_{ab}$. Note that we use the metricity condition, i.e. the metric in unaffected 
by the covariant derivatives.
The sum over all possible symmetric permutations of the indices is denoted
by the round brackets in the index, e.g. $t_{(ab)}=\frac12(t_{ab}+t_{ba})$. Note, due to the
metric-affinity of the Levi-Civita connection, $\nabla g=0$,
there are no terms coming from the metric in \eqref{eq:44-1}. An important subtlety arises due to the fact that the covariant derivatives do not commute which follows from the definition of the Riemann tensor $R(X,Y)Z$
\begin{equation}
    {R^a}_{bcd}X^bY^cZ^d=(\nabla_d\nabla_cZ^a-\nabla_c\nabla_dZ^a)X^cY^d
\end{equation}
for arbitrary vector fields $X$, $Y$, and $Z$.

Next, we define the complete set of solutions by
$f^n_{ab}$ and $d^n_{ab}$ with respect to the symplectic product $(\cdot,\cdot)$ as
\begin{align}
\left(f^n_{cd},f^m_{ab}\right)&=\frac{1}{2}\left(g_{ac}g_{bd}+
g_{bc}g_{ad}\right)\delta_{nm},\label{3.100}\\
\left((f^n_{cd})^{*},(f^m_{ab})^*\right)&=-\frac{1}{2}\left(g_{ac}g_{bd}+g_{bc}g_{ac}\right)\delta_{nm},\label{3.101}\\
\left(f^n_{cd},(f^m_{ab})^*\right)&=0\label{eq:5-1}
\end{align}
where $g_{ab}$ represents the background metric;
similar expressions hold for $d^n_{cd}$. Those mode functions can be used to 
expand the field into creation and annihilation operators. To do this, we define in- 
and out-vacuum states. In order to express the fields asymptotically as free states, we impose restrictions on the background space-time, i.e. the space-time must support 
asymptotically free states which are effectively not affected by the curvature. These set-ups are 
realized in asymptotically flat or asymptotically simple space-times.

From now on we specialize to space-times which support 
asymptotic theory such that the LSZ reduction can be performed \cite{fried92}.
Using this, we represent the interacting fields --using asymptotical freedom in the asymptotic
regions-- $h_{ab}^{\mathrm{in}}$ and $h_{ab}^{\mathrm{out}}$
in a complete basis 
\begin{equation}
\begin{aligned}
h_{ab}^{\mathrm{in}}(x) & = \sum_n\left[a_{n}^{\mathrm{in}}f^n_{ab}(x)+(a_{n}^{\mathrm{in}})^{\dagger}(f^n_{ab}(x))^{*}\right]\\
& = \sum_n\left[b_{n}^{\mathrm{in}}d^n_{ab}(x)+(b_{n}^{\mathrm{in}})^{\dagger}(d^n_{ab}(x))^{*}\right]\label{eq:1b}\\
h_{ab}^{\mathrm{out}}(x) & = \sum_n\left[a_{n}^{\mathrm{out}}f^n_{ab}(x)+(a_{n}^{\mathrm{out}})^{\dagger}(f^n_{ab}(x))^{*}\right]\\
& = \sum_n\left[b_{n}^{\mathrm{\mathrm{out}}}d^n_{ab}(x)+(b_{n}^{\mathrm{\mathrm{out}}})^{\dagger}(d^n_{ab}(x))^{*}\right]
\end{aligned}
\end{equation}
Given the Bogolubov transformation, the fields can be either expressed by the ladder operators 
$a_n$ or $b_n$ which are the usual annihilation operators with respect to the individual
vacuum and its corresponding eigenbases $f_{ab}^n$ or $d_{ab}^n$. We wanted to mention that 
other dependencies, e.g. momentum, are stored in the meta-label $n$.

Quantization of the system requires to construct {}``in'' and {}``out'' Fock spaces.
In static space-times, there will be a natural definition
of positive frequency modes due to the existence of a global timelike Killing vector field. 
The positive frequency solutions $f^{n}_{ab}$ fulfill the equation of motion
and can be used to give a natural definiton of the vacuum \cite{bir80}.

In curved space-time, this basis is not unique 
because Poincar\'e symmetry is not necessarily present. In fact, we have 
infinitely many nonequivalent vacua, thus, the decomposition into 
positive and negative frequency modes is not unique \cite{ash75}. A second peculiarity is given by 
asymptotic theory, that is to say, fields become effectively free 
$h_{ab}(t,x)\to h_{ab}^{\rm in}(t,x)$
in the limit $t\rightarrow-\infty$ and $h_{ab}(t,x)\to h_{ab}^{\rm out}(t,x)$
in the limit $t\rightarrow+\infty$. Additionally, we assume they
can be described as momentum eigenstates \cite{bir82}
such that we can substitute the creation or annihilation operators
by expressions in terms of the field operator. This presumes the existence of asymptotic particle states or, in cosmological settings, an adiabatic particle interpretation in the WKB regime.
Suppose the modes $f^n_{ab}$ are positive frequency 
as $t\rightarrow-\infty$ and $d^n_{ab}$ as 
$t\rightarrow+\infty$. Thus, the modes $f^n_{ab}$ and 
$(f^n_{ab})^{*}$ can be related with  $d^n_{ab}$ and 
$(d^n_{ab})^{*}$ by a Bogolubov transformation 
\begin{align}
f^n_{ab}=\sum_{m}\left(\alpha_{nm}^{*}d^m_{ab}+\beta_{nm}(d^m_{ab})^{*}\right)\label{eq:14}\\
d^n_{ab}=\sum_{m}\left(\alpha_{nm}f^m_{ab}+\beta_{nm}(f^m_{ab})^{*}\right)\label{eq:15}
\end{align}
where the Bogoliubov coefficients $\alpha_{nm}$ and $\beta_{nm}$
are time-independent complex matrices
which obey
\begin{equation}
\sum_{r}\left(\alpha_{mr}\alpha_{nr}^{\ast}-\beta_{mr}\beta_{nr}^{\ast}\right)=\delta_{mn}\label{eq:16}
\end{equation}
Our aim is evaluating the amplitude 
$\langle\mathrm{out}, n\mathcal{j}m, \mathrm{in}\rangle$. Note, that due to the non-trivial vacuum 
structure in curved space-times even the vacuum persistency amplitude
$\langle\mathrm{out},n\mathcal{j}m,\mathrm{ in}\rangle=\langle\mathrm{out}, n\mathcal{j}\mathcal S\mathcal{j}m, \mathrm{in}\rangle$ incorporates 
a scattering operator in order to translate between the states. The $\mathcal S$-operator 
contains vacuum expectation values of time ordered fields operators \cite{bir80,bir82}
\begin{equation}
\begin{aligned}
\mathcal{h}&\mathrm{out},n^{(s)}\mathcal{j}\mathcal S\mathcal{j}m^{(r)},\mathrm{in}\mathcal{i}=\\
&\sum_{\tau}i^{\frac{r-l}{2}}\prod_{p=1}^k\mathscr{O}_{z_{\tau(p)}}\prod_{q=k+1}^l \alpha^{-1}_{m_{\tau(q)}n_{\tau(q)}}\prod_{j=l+1}^{r-1}\Lambda_{\tau(j)\tau(j+1)} \\
&\times\sum_{\sigma}i^{\frac{s-l-w+k}{2}}\prod_{\iota=1}^w\mathscr{Q}_{z_{\sigma(\iota)}}\prod_{\upsilon=w+1}^{s-l+k-1}V_{\sigma(\upsilon)\sigma(\upsilon+1)}\\
&\times\mathcal{h}\mathrm{out}\mathcal{j}\mathscr{T} \{h_{ab}(x_{\sigma(\iota)})h_{ab}(x_{\tau(p)})\mathcal{j}\mathrm{in}\mathcal{i}\label{3.27}
\end{aligned}
\end{equation}
where we have made several definitions: The integral operators $\mathscr{O}_{x_{i}}$ 
and $\mathscr{Q}_{x_{i}}$ are with respect to the
two different mode functions $d_n^{ab}$ and $f_l^{ab}$
\begin{eqnarray}
\mathscr{O}_{x_{i}}&\equiv i\underset{l}{\sum}\alpha_{m_{i}l}^{-1}\int\left[d_l^{cd}(x_{i})\left({\mathscr{D}^{ab}}_{cd}\right)_{x_{i}}\right]\sqrt{-g_{x_{i}}}{\rm d}^{4}x_{i}\mbox{\,\, }\label{3.109}\\
\mathscr{Q}_{x_{i}}&\equiv i\underset{l}{\sum}\alpha_{ln_{i}}^{-1}\int\left[f_l^{cd}(x_{i})\left({\mathscr{D}^{ab}}_{cd}\right)_{x_{i}}\right]\sqrt{-g_{x_{i}}}{\rm d}^{4}x_{i}\mbox{\,\, }\label{3.1099}
\end{eqnarray}
where the permutations of the indices are given by $\tau$ and $\sigma$, while the coefficients of the matrices $\boldsymbol{\Lambda}$ and $\boldsymbol{V}$ are
determined by the Bogolubov coefficients
\begin{eqnarray}
\Lambda_{ij}&\equiv-i\underset{n}{\sum}\alpha_{m_{i}n}^{-1}\beta_{nm_{j}}\label{3.110}\\
V_{ij}&\equiv i\underset{n}{\sum}\beta^{\ast}_{m_{j}n}\alpha_{nm_{i}}^{-1}\label{3.1100}
\end{eqnarray}
When we factor out the polarization in $d_n^{ab}$
to get $d_n^{ab}=d_n\varepsilon^{ab}$,
where the index $n$ no longer encodes the polarization of the fields.
Our next step is to substitute the polarization tensor with the momentum,
therefore we split $\varepsilon^{ab}$ into its vector parts 
and perform the high-energy limit. Thus, we can replace $\varepsilon^{ab}$
by the longitudinal polarized part $\varepsilon_{L}^{ab}$
which in turn yields\footnote{The decoupling limit in curved space-times is rather local than global; additionally one may need to consider low curvature, i.e. $R\ll E^2$ such that some operators become curvature suppressed as $R/m^2$.}
\begin{equation}
\varepsilon_{L}^{ab}=\frac{k^{a}k^{b}}{m^{2}}+\mathcal{O}\left(\frac{m}{E}\right)
\end{equation}
Let us proceed by using only one momentum at a time. Contrary to the flat metric
we substitute $k_a$ with covariant derivatives to be consistent with the equations of motions in curved space-time.
In order to finally prove the GBET we get the Goldstone
fields and relate them to the spin-2 tensor field $h_{ab}(x)$.
The underlying symmetry provides the Gupta-Bleuler condition 
and the gauge-fixing
\begin{equation}
\nabla^{a}h_{ab}=\nabla_{b}h+\lambda mA_{b}.\label{gb}
\end{equation} 
After inserting we get a part of the Lagrange density only depending on the Goldstone vector, that is,
$-\mathcal{L}_{A} = A_a\mathcal{D}^{ab}A_b$, 
with $\mathcal{D}^{ab}$ being the differential operator for $A_a$
\begin{equation}
\mathcal{D}_{ab}=\left[\frac{1}{4}g_{ab}\Box -\frac{1}{8}\nabla_{b}\nabla_{a}-\frac{1}{8}\nabla_{a}\nabla_{b}+\frac{3}{8}R_{ab}+\frac{\lambda}{2}m^{2}g_{ab}\right]
\end{equation}
After using the gauge freedom, we find the following condition to
hold
\begin{equation}
\nabla^{b}{\mathscr{D}^{cd}}_{ba}h_{cd}=-m{\mathcal{D}^{b}}_{a}A_{b}.
\end{equation}
where the explicit calculations are found in Appendix \ref{appb}. Thus, we could formulate the first step of the GBET on curved space-times, i.e. a reduction 
to Goldstone vectors
\begin{equation}
\begin{aligned}
\varepsilon&^{b}k^{a}\mathcal{M}_{ab}(\psi_{\rm in}+h_{ab}\rightarrow \psi_{\rm out}) \\
 &=  C_{i}\langle\mathrm{out},n^{(s)}\mathcal{j}m\!\!\int \!\!\left[\varepsilon^{b}\mathcal{D}^{a}_{\,\,\,b}A_{a}\right]\sqrt{-g}\,\mathrm{d}^{4}x\mathcal{j}m^{(r)}-m_{i},\mathrm{in}\rangle\label{eq:45}
\end{aligned}
\end{equation}
for the rest matrix element. In the high-energy limit, we use the condition
that the longitudinal polarization dominates and we find the desired 
formulation of the GBET 
\begin{equation}
\varepsilon^{b}\varepsilon_{L}^{a}\mathcal{M}_{ab}(\psi_{\rm in}+h_{ab}\rightarrow \psi_{\rm out})  =  C\varepsilon^{b}\mathcal{M}_{b}(\psi_{\rm in}+A_{b}\rightarrow \psi_{\rm out})\label{epsi}
\end{equation}
In the high-energy limit the theory is - according to
\eqref{epsi} - accurately described by a spin-1 particle, that is the Goldstone vector field.
In this regime our theory reduces effectively to
\begin{eqnarray}
\mathcal{L}_{\tilde{A}} & = & -\frac{1}{8}F_{ab}F^{ab}+\frac{1}{2}\tilde{A}^{a}R_{ab}\tilde{A}^{b}+\frac{1}{2}\lambda m^{2}\tilde{A}_{a}\tilde{A}^{a}\label{eq:47}
\end{eqnarray}
Again, we make use of the St\"uckelberg formalism and 
introduce an additional scalar degree of freedom
\begin{equation}
\tilde{A}_{a}\rightarrow A_{a}+\frac{\nabla_{a}\phi}{m}\label{zwei}.
\end{equation}
In order to extract the kinetic operators, we need to
diagonalize \eqref{eq:47}. Hence, we choose for the gauge-fixing
\begin{equation}
\mathcal{L}_{\rm gf}=\frac{1}{2\zeta}\left[\nabla^{b}\left(\frac{A^{a}R_{ab}}{m}+\lambda mA_{b}\right)+\zeta m\phi\right]^{2},\label{eq:49}
\end{equation}
where the gauge parameter $\zeta$ denotes the gauge coupling. Our effective theory $ \mathcal{L}_{A}+\mathcal{L}_{\rm gf}$ including the Goldstone scalar is then described by
\begin{align}
\mathcal{L}_A&+\mathcal{L}_\phi=A^{b}\left[\frac{1}{4}\Box g_{b}^{a}-\frac{1}{8}\nabla_{b}\nabla^{a}-\frac{1}{8}\nabla^{a}\nabla_{b}+\frac{3}{8}R_{b}^{a}\right.\nonumber\\
 &\left.\hspace{-0.3cm}+\frac{\lambda m^{2}g_{b}^{a}}{2}-\frac{\left(R_{c}^{a}+\lambda m^2g_{c}^{a}\right)\nabla^{c}\nabla^{d}\left(R_{bd}+\lambda m^2g_{bd}\right)}{2\zeta m^2}\right]A_{a}\nonumber\\
  & \hspace{-0.3cm}+\phi\left[-\frac{1}{2m^{2}}\nabla^{a}\left(R_{ac}\nabla^{c}\right)-\frac{\lambda}{2}\Box+\frac{\zeta m^2}{4}\right]\phi\label{eq:50}
\end{align}
Consider the matrix element for the massive spin-1 particle 
- but this time in the high-energy limit. For convenience, we call the expression in brackets $\bar{\mathcal{D}}_{ab}$. With \eqref{eq:49}, we get
\begin{equation}
\nabla^{b}{\bar{\mathcal{D}}^{a}}_{b}A_{a} =-\frac{1}{4}\zeta m^3\phi+\frac{1}{2m}\nabla^{b}\left[R_{bc}\nabla^{c}\phi\right]+\frac{1}{2}\lambda m\Box\phi\label{eq:52}
\end{equation}
Comparison to the kinetic operator for the scalar field \eqref{eq:50} shows that we have picked up an extra term that is proportional to the Ricci curvature tensor. This term inflicts a direct modification of the kinetic operator, such that $\bar\Box=\mathfrak{g}^{ab}\nabla_a\nabla_b$ where the effective metric $\mathfrak{g}$ contains a Ricci tensor modification
\begin{equation}\label{effectivemetrik}
    \mathfrak{g}_{ab}=g_{ab}+\frac{1}{\lambda m^2}R_{ab}, 
\end{equation}
which comes with a mass suppression. This feature is important, because it will introduce a limit to the validity of the Stückelberg procedure within curved space-times. Additionally, this extra-term highlights the difference between high-energy limit and high-curvature region. In curved space-times these two limits do not necessarily coincide which is due to the fact that there exists no Fourier conjugation between configuration and momentum space. In this sense, we need to be careful with both limits and the additional Ricci-term keeps track of this, as we will see for the FLRW space-time in the next section.

We would expect such a limit to be prevalent, since we work in a linearized theory with weakly coupled degrees of freedom. However, as effective field theory studies of quantum fields on curved backgrounds show, even the free field description can, in certain setups, reach the boundaries of validity of the description no matter the initial conditions. This may happen when, due to curvature effects, the field experiences space-time friction that renders the effective description insignificant \cite{PhysRevD.111.025019}.

On the other hand, the effective metric does not necessarily need to support a well-defined kinetic term since, in principle, the sign in $\mathfrak{g}$ is not a priori fixed, leading to a possibly wrong sign in the kinetic term which is associated with an instability in the system. From this perspective, the methodology we used serves as a sensitive indicator to study these bounds explicitly. 

As such, this term is potentially harmful, as it may trigger several instabilities. Since the term itself remains second order in derivatives, Ostrogradskiy instabilities are not expected to occur. Although the LSZ reduction formalism has reduced the linearized spin-2 tensor field to only the healthy Goldstone scalar degree of freedom (the sixth ghost degree of freedom was excluded), the modified kinetic operator could still cause instabilities. 

Hence, constructing the kinetic operator $D$ with $\mathfrak g$ gives the effective differential operator $\bar D$. Nevertheless, due to the locality of the decoupling limit in curved space-times, such operators could also be considered to decouple in the high energy limit as long as $R\ll E^2$ and, thus, these terms become curvature suppressed by $R/m^2$.
In this sense, we find 
\begin{equation}
\nabla_{b}\bar{\mathcal{D}}^{ab}A_{a}=-m\bar D\phi.
\end{equation}
Putting everything together, it turns out that through successively
 reducing the spin-2 tensor field we can formulate the Goldstone boson 
equivalence theorem for deformed spin-2 tensors on curved space-times
 \begin{align}
\varepsilon^{cd}_{L}&\mathcal{M}_{cd}(\psi_{\rm in}+h_{ab}\rightarrow \psi_{\rm out}) \label{eq:53}\\
&=  -C\mathcal{M}(\psi_{\rm in}+\phi\rightarrow \psi_{\rm out})\left[1+\mathcal{O}\left(\frac{E}{m}\right)+\mathcal{O}\left(\frac{R}{m^2}\right)\right]\nonumber
\end{align}
which concludes that the Goldstone boson equivalence
theorem is valid for high enough energies and sufficiently large curvature radii.

\section{Cosmological Space-time}

The Friedmann-Lema\^itre-Robertson-Walker (FLRW) space-time is, in cosmological time, given by
\begin{equation}
    g=-{\rm d}t\otimes{\rm d}t+a^2(t)\left(\frac{{\rm d}r\otimes{\rm d}r}{1-\kappa r^2}+r^2{\rm d}\mathbb{S}_2\right),
\end{equation}
where $a(t)$ denotes the scale factor and d$\mathbb{S}_2$ the line-element of the 2-sphere. The sectional curvature is determined through $\kappa$. It captures three cases: a flat geometry (flat universe) with $\kappa=0$, a spherical geometry (closed universe) for $\kappa=1$, and a hyperbolic geometry for $\kappa=-1$ (open universe). 

Regarding particular examples, we study the radiation dominated universe, i.e. $a(t)=\sqrt{t}$ and the de Sitter scenario $a(t)=e^{\mathcal{H}_0t}$ with $\mathcal{H}_0$ a constant. While radiation-dominated universes describe highly energetic regions such as right after the big bang, while the de Sitter phase captures inflationary scenarios. 

We would like to understand the impact of the effective kinetic operator on an FLRW background. For the sake of simplicity, we choose the flat case and work in Cartesian coordinates. In this geometry the Ricci tensor becomes diagonal with non-zero components
\begin{equation}
R^{tt}=-3(\dot{\mathcal{H}}+\mathcal{H}^2),\;R^{ii}=(\dot{\mathcal{H}}+3\mathcal{H})g^{ii},\;\forall i\in\{x,y,z\},
\end{equation}
where we defined the Hubble parameter $\mathcal{H}(t)=\frac{\dot a(t)}{a(t)}$. Using \eqref{effectivemetrik} and the $\phi$-part of the Lagrange density \eqref{eq:50}
\begin{equation}
    \mathcal{L}\supset\mathcal{L}_\phi=-\frac12\phi\left[\mathfrak{g}^{ab}\nabla_a\nabla_b-\frac{\zeta m^2}{2}\right]\phi\,,
\end{equation}
we discover several potential cases for instabilities. They are related to the components of $\mathfrak{g}$. The first instability would be a ghost instability originating from the $tt$-component. This pathology is prevented as long as
\begin{equation}\label{geistbed}
0<1+\frac{3(\dot{\mathcal{H}}+\mathcal{H}^2)}{\lambda m^2}.
\end{equation}
The spatial coefficients could generate a similar instability due to a sign change. Hence, whenever
\begin{equation}\label{gradienteninst}
0<1+\frac{\dot{\mathcal{H}}+3\mathcal{H}^2}{\lambda m^2},
\end{equation}
the setup remains healthy. A violation of \eqref{gradienteninst} would result in an exponential growth for large-$k$ modes. The third instability is connected to the propagation speed of fluctuations. A modified propagator will trickle down to the dispersion relation
\begin{equation}
    \omega^2=c_{\rm s}^2\frac{k^2}{a^2}+\frac{\zeta m^2}{2},
\end{equation}
where we defined the effective signal speed $c_{\rm s}$ as
\begin{equation}
    c_{\rm s}^2=\frac{\lambda m^2+\dot{\mathcal{H}}+3\mathcal{H}^2}{\lambda m^2+3\dot{\mathcal{H}}+3\mathcal{H}^2}.
\end{equation}
It is clear that whenever $c_{\rm s}^2\le0$ we develop a gradient instability that leads to an exponentially growing mode while the condition $c_{\rm s}>1$ would violate the rules of information processing due to superluminal propagation speed. Regarding the latter, we would retrieve a condition
\begin{equation}
    2\dot{\mathcal{H}}<3\mathcal{H}^2\quad\Leftrightarrow\quad2\frac{\ddot{a}}{a}<3\left(\frac{\dot{a}}{a}\right)^2.
\end{equation}
To appreciate the information coming from these inequalities, we consider two examples of specific matter content: a cosmological constant (de Sitter universe) and the radiation filled universe. 

\subsection{De Sitter universe}

Starting with the first example, $a(t)=e^{\mathcal{H}_0t}$, where $\mathcal{H}=\mathcal{H}_0$ and $\dot{\mathcal{H}}=0$. In this case, most conditions are trivially obeyed (at least under meaningful assumptions, e.g. $\mathcal{H}^2>0$ and $m^2>0$). The bounds from both, the temporal as well as the spatial gradient fulfill $\lambda m^2+3\mathcal{H}^2>0$, therefore, there exists neither a ghost nor a gradient instability. Since, $\dot{\mathcal{H}}\equiv0$ everywhere, $c_{\rm s}=1$ for de Sitter patches. This resonates with the simulation based analysis in \cite{PhysRevD.111.025019}.

\subsection{Radiation-filled universe}

The radiation-filled universe, that is, $a(t)=\sqrt{t}$, offers a much richer phenomenology with the big bang singularity at $t=0$. As a consequence, the condition for a lack of ghosts is not necessarily fulfilled because $R_{00}=-3\frac{{\ddot a}}{a}=-\frac{3}{4t^2}$ is negative definite for $t>0$. Hence, \eqref{geistbed} is violated whenever
\begin{equation}
    t<\sqrt{\frac{3}{4m^2\lambda}}
\end{equation}
because the sign of the kinetic term changes. To turn it around, the Goldstone description becomes predictive, after a certain distance from the big bang. In regions, where the curvature radius becomes small with respect to the mass of the scalar, we see that the GBET breaks down because it becomes a ghost degree of freedom. The sign change only occurs in the $tt$-component whilst the spatial components remain positive definite and therefore \eqref{gradienteninst} is always fulfilled such that no gradient instability is triggered. 

The late time consistency is maintained as well for the signal propagation speed, which reads
\begin{equation}
    c_{\rm s}=\frac{4\lambda m^2t^2+1}{4\lambda m^2t^2-3}
\end{equation}
which for large times becomes approximately one, while small times create huge digressions from the speed of light, such that for a similar range $t\sim\frac1m$, the description shows pathological features. However, this deformation of the signal speed is parametrically small, i.e. $c_s^2-1\sim\frac{\mathcal{H}^2}{m^2}$ which is within the errors induced by the GBET itself, that is $\mathcal{O}(\frac{R}{m^2})$.

\subsection{Open and closed universes}

Finally, we classify our findings with respect to the strong coupling regime with cutoff $\Lambda_{3}\sim(m^2m_{\rm pl})^{1/3}$, where $m_{\rm pl}$ denotes the Planck mass \cite{derham}. Effectively, the curvature contributions admit a mass suppression of the kind $\frac{R}{m^2}$. The spin-0 mode enters the strong coupling regime once $R\sim\mathcal{H}^2\gtrsim m^2$. For radiation dominated universes, a majority of the modes will not obey this condition for small times \cite{PhysRevD.111.025019}, hence, the effective field theory description will break down long before the ghost regime will be reached. When analyzing the effective field theory bound for de Sitter, we consider the Higuchi bound \cite{higuchi1987forbidden}, $m^2\ge2\mathcal{H}_0^2$, which limits our description beyond the limits set by the effective metric.

Let us comment on the possibilities to incorporate the other sectional curvatures $\kappa=\pm1$ as well. The issue is that we cannot rely on the Fourier transformation, since the harmonic functions $u_p$ in spaces with non-zero sectional curvature are different; hence, we must use the harmonic equation $\Delta u_p=\lambda_pu_p$ with $p\in\{k,n,q\}$ such that all cases are accounted for.
The precise spectrum depends on the spatial curvature:
\begin{align}
\kappa=0\quad\Rightarrow&\quad \lambda_k=k^2, \hspace{3.7em} k\in\mathbb{R}^{+},\\
\kappa=+1\quad\Rightarrow&\quad \lambda_n=n(n+2), \quad n\in\mathbb{N}_0,\\
\kappa=-1\quad\Rightarrow&\quad \lambda_q=q^2+1, \hspace{2.1em} q\in\mathbb{R}_>.
\end{align}
The quantity $\sqrt{\lambda_p}$ therefore plays the role of the comoving momentum magnitude, while the associated physical momentum would be $p=\sqrt{\lambda_p}/a(t)$. In the short-wavelength regime $p\ll\mathcal H$, the harmonics admit a local WKB interpretation and derivative operators may be replaced parametrically by powers of $\sqrt{\lambda_p}$, reproducing the standard flat-space momentum correspondence. 

It is important to stress that the harmonics are not truly eigenfunctions of the single derivative operator in the same sense as plane waves are. Rather, they diagonalize the Laplace-Beltrami operator, which is the covariant second-order spatial operator compatible with homogeneity and isotropy. Consequently, the identification $\sqrt{\lambda_p}\sim |\boldsymbol{k}|$ should be understood in terms of spectral theory: it characterizes the inverse wavelength or gradient scale of the mode, not the eigenvalue of an individual covariant derivative. 

In this sense, due to the essential self-adjointness of $\Delta$ one may take the formal square root of the harmonic equation to associate a momentum with the single derivative, i.e. $\sqrt{\Delta}=\sqrt{\lambda_p}$ then implies $|\nabla|\rightarrow \sqrt{\lambda_p}$ mode by mode. 

The physical justification originates from the short-wavelength or Wentzel-Kramers-Brillouin (WKB) regime. We assume that this approximatiom holds such that we are able to define instantaneous vacuum states \cite{ag15}. Upon a formal harmonic decomposition of the field,
\begin{equation}
\phi(x)=\int\mbox{d}^3p\,\varphi_p(t)u_p(\boldsymbol x)
\end{equation}
the Klein-Gordon equations transforms into a free massive scalar mode obeys the wave equation
\begin{equation}
\ddot{\varphi}_{p}
+3\mathcal H\dot{\varphi}_{p}
+
\left(
m^2+\frac{\lambda_p}{a^2(t)}
\right)\varphi_{p}=0.
\end{equation}
Assuming that the mode frequency varies slowly compared with the oscillation time scale, one can use the ansatz
\begin{equation}
\varphi_{p}(t)
=
A(t)
\exp\!\left(-i\int^t_0 \omega_p(\tau){\rm d}\tau\right).
\end{equation}
where $A(t)$ denotes the amplitude here. At zeroth order in adiabaticity \cite{parker1974adiabatic,anderson1987adiabatic} yields
\begin{equation}
\omega^{(0)}_p(t)\simeq \sqrt{m^2+\frac{\lambda_p}{a(t)^2}}.
\end{equation}
Thus, in the adiabatic expansion and within the WKB approximation, the harmonic mode behaves locally as a field on a flat background with physical momentum $p=\sqrt{\lambda_p}/a(t)$. Note, the validity of the WKB expansion is determined through $|\dot\omega_p|\ll|\omega^2_p|$ which for high-energetic modes reduces parametrically to
\begin{equation}
\mathcal H\ll\frac{\sqrt{\lambda_p}}{a(t)}.
\end{equation}
In that limit, the mode probes only a local region of space-time, for which curvature effects are negligible.

This observation is particularly relevant for the GBET because what is referred to as high energies should in fact be small distances in curved space-times. Hence, the theorem should be understood as holding in the joint high-energy and short-wavelength regime for which longitudinal polarizations reduce locally to their Goldstone counterparts up to corrections. While the harmonic decomposition provides the natural curved-space replacement of Fourier momentum methods, the WKB limit supplies the dynamical justification for their use.

In fact, the harmonics provide then a spectral replacement, that is valid mode-by-mode,
and becomes asymptotically exact in the UV (or WKB) limit. In this sense, the GBET for generic FLRW space-time would read
\begin{align}
    \varepsilon^{cd}&\mathcal{M}_{cd}(\psi_{\rm in}+h_{ab}\to\psi_{\rm out})=C\mathcal{M}(\psi_{\rm in}+\phi\to\psi_{\rm out})\nonumber\\
&\hspace{2em}\times\left[1+\mathcal{O}\left(\frac{ma(t)}{\sqrt{\lambda_p}}\right)+\mathcal{O}\left(\frac{\mathcal{H}(t)a(t)}{\sqrt{\lambda_p}}\right)\right],
\end{align}
where we expressed the physical quantities through the spectral values of the harmonics.

\section{Conclusion}

We showed that the Goldstone boson equivalence theorem can be formulated on a certain class of flt space-times. Within the high energy regime, even a spin-2 theory is well described by the longitudinal scalar degree of
freedom. Our analysis unveiled additionally that the GBET can only be trusted as long as the curvature radius is large compared to the energy of the field. Once this threshold is exceeded, the Goldstone degree of freedom may cause instabilities signaling a failure of the description. We would interpret this as the need for incorporating backreaction. Altogether this shows that the GBET is a remarkably powerful tool and provides robust, qualitative and quantitative estimates for an effective high-energy theory. Colloquially speaking, the Goldstone boson equivalence theorem (GBET) is a proxy for the full theory at energy scales that are much higher than the particle's mass and curvature radii that are larger then the particles energy.

In this article we have used a versatile method to derive the GBET based on the Lehmann-Symanzik-Zimmermann (LSZ) reduction formalism. In particular, we showed that the formalism extends the validity of the GBET for the subclass of cosmological space-times like the flat Friedmann-Lema\^itre-Robertson-Walker background up to the case of spin-2 tensor fields. The generalization of the GBET to spin-2 particles justifies that there shall exist a GBET for general higher spin theories as well, i.e. at high energies massive (infrared deformed) $(r,s)$-tensor
fields shall be well described by the longitudinal scalar degree of freedom which is responsible for
strong-coupling effects, decoupling limits, and potential instabilities. In this sense, the significance of the result lies not primarily in the computation of graviton scattering observables, but rather in the insights into the dynamics of the longitudinal sector.

Moreover, the outlined treatment does not require an underlying Abelian gauge
symmetry. For the standard model an analog technique can be applied \cite{hor97}.
Instead of using the Gupta-Bleuler condition, non-Abelian theories
require BRST charges and the addition of a Fadeev-Popov Lagrange 
density in order to cancel anomalies. Besides this, the GBET works 
similar to the procedure outlined in this article and can be implemented
in curved space-times too.

However, we saw that the formalism propagates potential instabilities coming from a curvature contribution in the kinetic operator. For the FLRW space-times, we investigated how the additional term inflicts instabilities and compared them with the effective field theory bound. It turned out that for a de Sitter (inflationary) universe, there are no potential threats, while a radiation dominated universe may admit potential instabilities. However, those pathologies were only triggered in the high curvature regime, i.e. close to the singularity, and hence, they lie potentially outside the effective field theoretic description. In this sense, the GBET offers a probing mechanism for the range of validity of the Goldstone description on curved spacetimes. 

Here, we made use of the fact that we can replace the momentum by a derivative which is a consequence of the Ward identity. Since our analysis has been performed in spatially flat space-time, we were able to use a replacement à la Weinberg where we traded the partial derivative that comes from the Fourier inversion of the polarization by a covariant derivative. The idea works also for FLRW space-times with different sectional curvature, that is, in situations where the spatial slices are spheres or hyperbolic spaces. By using the spectrum of the Laplace-Beltrami operator, we can associate the eigenvalues with a comoving momentum in the high-energy limit (or better in the small-distance limit).

However, in general curved space-times, this method cannot be assumed to go through so easily. In fact, one would probably work is to identify the vector field of the momentum and then use the harmonics and the related eigenvalue equation as we have done for the non-flat FLRW space-times. However, another complication impose the Ward identities and therefore the LSZ reduction which in generally curved space-times may not be straightforward \cite{mandal2024ward,de2025gravitational,friedman1992unitarity}. Albeit being true for highly symmetric space-times like FLRW, we forgo the claim that we have no loss of generality.

The present construction may be generalized to a broader class of backgrounds. Of particular interest would be Einstein backgrounds which allow a consistent formulation of massive spin-2 theories \cite{buchbinder2000equations}. In such an approach, the role of the momentum variable would be replaced by the spectral parameter of the relevant Laplace-Beltrami operator -- provided it admits an essentially self-adjoint extension --, with the associated momentum scale determined by the square root of its eigenvalues. However, a rigorous implementation requires a careful analysis of the spectral properties of the underlying differential operators. Since these issues are strongly geometry-dependent,
the question how our method can be adapted to more generic backgrounds will be a matter of further research.

\begin{acknowledgments}
Thank you to Sophia Zielinski and Stefan Hofmann for helpful discussions during the early stages of this investigation. This work was supported by the Basque Government Grant
\mbox{IT1628-22} and by the Grant PID2021-123226NB-I00 (funded by
MCIN/AEI/10.13039/501100011033 and by ``ERDF A way of making Europe''). \end{acknowledgments}

\appendix

\section{Explicit form of the Lagrange density}\label{appa}

The Lagrange density for a spin-2 tensor field on a curved background including the gauge-fixing condition is found to be 
\begin{eqnarray}
\mathcal{L} && = \! \mathcal{L}_{EH}-\frac{1}{2}h_{ab}T^{ab} 
+h^{ab}\!\!\left[-\frac{m^{2}}{4}\left(g_{a}^{c}g_{b}^{d}+g_{a}^{c}g_{b}^{c}-2g_{ab}g^{cd}\right)\right.\nonumber\\
&&\left.-\frac{1}{16\lambda}\left(\nabla_{a}\nabla^{c}g_{b}^{d}+\nabla_{b}\nabla^{c}g_{a}^{d}+\nabla_{a}\nabla^{d}g_{b}^{c}+\nabla_{b}\nabla^{d}g_{a}^{c}\right)\right.\nonumber \\
& & -\frac{1}{16\lambda}\left(\nabla^{c}\nabla_{a}g_{b}^{d}+\nabla^{c}\nabla_{b}g_{a}^{d}+\nabla^{d}\nabla_{a}g_{b}^{c}+\nabla^{d}\nabla_{b}g_{a}^{c}\right.\\
&&\left.-R_{\,\,a}^{c}g_{b}^{d}-R_{\,\,b}^{c}g_{a}^{d}-R_{\,\,a}^{d}g_{b}^{c}-R_{\,\,b}^{d}g_{a}^{c}\right) +\frac{1}{8\lambda}\left(R_{b\,\,\,a}^{\,\,dc}+R_{b\,\,\,a}^{\,\,cd}\right)\nonumber \\
& & \left.+\frac{1}{4\lambda}\left(\nabla_{a}\nabla_{b}g^{cd}+\nabla_{b}\nabla_{a}g^{cd}+g_{ab}\nabla^{c}\nabla^{d}+g_{ab}\nabla^{d}\nabla^{c}\right)\right.\nonumber\\
&&\left.-\frac{1}{2\lambda}\!\!\left(g_{ab}\square g^{cd}\right)\!\right]\!\!h_{cd}\!-\!\frac{1}{8}F_{ab}F^{ab}\!+\!\frac{1}{2}R_{ab}A^{a}A^{b}\!+\!\frac{\lambda m^{2}}{2}A_{a}A^{a}\nonumber\label{eq:44-1}
\end{eqnarray}
where $\mathcal{L}_{\rm total}=\mathcal L + \mathcal{L}_{\rm gf}$ with the gauge-fixing term \eqref{gb}.

\section{Calculations in Flat Spacetime}\label{appb}

Our starting point of the calculation that is involved in the proof of the GBET for flat space-time will be the kinetic operator ${\mathscr{D}^{ab}}_{cd}$ in \eqref{2.81} which looks after adding the gauge-fixing term \eqref{2.78}
\begin{equation}
\begin{aligned}
&{\mathscr{D}^{ab}}_{cd}=\frac{1}{2}\left(\Box-m^2\right)\left(\frac{1}{2}{\eta^a}_c{\eta^{b}}_d+\frac{1}{2}{\eta^{a}}_d{\eta^{b}}_c-\eta_{cd}\eta^{ab}\right)\\
&+\frac12\left(1+\frac1\xi\right)(\eta^{ab}\partial_{c}\partial_{d}+\eta_{cd}\partial^{a}\partial^{b})-\frac{1}{2\xi}\eta^{ab}\Box\eta_{cd}\\
&-\frac{1}{4}\left(1+\frac1{2\xi}\right)\!\!\left({\eta^b}_d\partial^{a}\partial_{c}+{\eta^b}_c\partial_d\partial^{a}+{\eta^a}_d\partial^{b}\partial_c+{\eta^a}_c\partial^{b}\partial_d\right)\label{2.81}
\end{aligned}
\end{equation}
where we omitted the $B_a$-dependent term. By applying a partial derivative, the kinetic operator can be rewritten as
\begin{equation}
\begin{aligned}
&\partial_a{\mathscr{D}^{ab}}_{cd}=\frac{1}{2}\left(\Box-m^2\right)\left(\frac{1}{2}{\eta^{b}}_d\partial_c+\frac{1}{2}{\eta^{b}}_c\partial_d-\eta_{cd}\partial^{b}\right)\\
&+\frac12\left(1+\frac1\xi\right)(\partial^{b}\partial_{c}\partial_{d}+\eta_{cd}\Box\partial^{b})-\frac{1}{2\xi}\eta_{cd}\Box\partial^{b}\\
&-\frac{1}{4}\left(1+\frac1{2\xi}\right)\!\!\left({\eta^b}_d\Box\partial_{c}+{\eta^b}_c\Box\partial_d+\partial_d\partial^{b}\partial_c+\partial_c\partial^{b}\partial_d\right).\label{2.81}\\
\end{aligned}
\end{equation}
When applying this onto $h^{cd}$, we can extract the following result

\begin{equation}
\begin{aligned}
&\partial_a{\mathscr{D}^{ab}}_{cd}h^{cd}=\frac{1}{2}\left(\Box-m^2\right)\left(\partial_ch^{bc}-\partial^{b}h\right)\\
&+\frac12\left(1+\frac1\xi\right)(\partial^{b}\partial_{c}\partial_{d}h^{cd}+\Box\partial^{b}h)-\frac{1}{2\xi}\Box\partial^{b}h\\
&-\frac{1}{2}\left(1+\frac1{2\xi}\right)\!\!\left(\Box\partial_{c}h^{cb}+\partial_c\partial^{b}\partial_dh^{cd}\right).\label{2.81}\\
&=\left[(\Box-m^2)-\left(1-\frac1\xi\right)\Box\right](\partial_ch^{bc}-\partial^bh)\\
&=\left(\frac{\Box}{\xi}-m^2\right)\xi m \tilde B^b.
\end{aligned}
\end{equation}
In the last step, we used the gauge condition to replace the spin-2 field by its longitudinal, spin-1 Goldstone boson. 

\section{Calculations in Curved Spacetime}

To determine the explicit form of $\nabla^{a}{\mathscr{D}_{ab}}^{cd}h_{cd}$, we use the relation between $h_{ab}$ and
the Goldstone (St\"uckelberg) vector field $\ensuremath{\nabla^{a}h_{ab}=\nabla_{b}h+\lambda mA_{b}}$
we get
\begin{eqnarray}
\nabla^{a}{\mathscr{D}_{ab}}^{cd}h_{cd} & = & -\frac{1}{2}\lambda m^{3}A_{b} -\frac{1}{8\lambda}\left(\square\nabla_{b}h+\lambda m\square A_{b}\right.\nonumber\\
&&+\underset{1.}{\underline{\nabla^{a}\nabla_{b}\nabla^{c}h_{ca}}}+\underset{2.}{\underline{\nabla^{a}\nabla^{c}\nabla_{a}h_{cb}}}-4\nabla_{b}\square h\nonumber \\
& &+\underset{3.}{\underline{\nabla^{a}\nabla^{c}\nabla_{b}h_{ca}}} -\left(\nabla^{a}R_{\,\,a}^{c}\right)h_{cb}-R_{\,\,a}^{c}\nabla^{a}h_{cb}\nonumber \\
& &-\left(\nabla^{a}R_{\,\,b}^{c}\right)h_{ca} -R_{\,\,b}^{c}\nabla^{a}h_{ca}-2\square\nabla_{b}h\nonumber \\
& & -2\left(\nabla^{a}R_{b\,\,\,a}^{\,\,dc}\right)h_{cd}-2R_{b\,\,\,a}^{\,\,dc}\nabla^{a}h_{cd}\nonumber\\
&&-2\underset{4.}{\underline{\nabla^{a}\nabla_{b}\nabla_{a}h}}-\left.4\nabla_{b}\square h-4\lambda m\nabla_{b}\nabla^a A_a\right)\nonumber
\end{eqnarray}
We provide the results for the underlined terms below.
\begin{enumerate}
\item  
\begin{eqnarray*}
\nabla^{a}\nabla_{b}\nabla^{c}h_{ca} & = & \nabla_{b}\nabla_{a}\nabla_{c}h^{ca}+R_{ab}^{\,\,\,ad}\nabla^{c}h_{cd}\\
& = & \nabla_{b}\square h+\lambda m\nabla_{b}\nabla A+R_{b}^{\,\,d}\nabla_{d}h+\lambda mR_{b}^{\,\,d}A_{d}
\end{eqnarray*}
\item  
\begin{eqnarray*}
\nabla^{a}\nabla^{c}\nabla_{a}h_{cb}
& = & \square\nabla^{c}h_{cb}+\left(\nabla_{a}R^{da}\right)h_{db}+R^{da}\nabla_{a}h_{db}\\
&&+\left(\nabla_{a}R_{b}^{\,\,dca}\right)h_{cd}+R_{b}^{\,\,dca}\nabla_{a}h_{cd}\\
& = & \square\nabla_{b}h+\lambda m\square A_{b}+\left(\nabla_{a}R^{da}\right)h_{db}\\
&&+R^{da}\nabla_{a}h_{db}+\left(\nabla_{a}R_{b}^{\,\,dca}\right)h_{cd}+R_{b}^{\,\,dca}\nabla_{a}h_{cd}
\end{eqnarray*}
\item  
\begin{eqnarray*}
\nabla^{a}\nabla^{c}\nabla_{b}h_{ca} & = & \nabla_{a}\nabla_{b}\nabla_{c}h^{ca}+\left(\nabla_{a}R_{db}\right)h^{da}+R_{db}\nabla_{a}h^{da}\\
&&+\left(\nabla_{a}R_{\,\,dcb}^{a}\right)h^{cd}+R_{\,\,dcb}^{a}\nabla_{a}h^{cd}\\
& = & \nabla_{b}\nabla_{a}\nabla_{c}h^{ca}+R_{bd}\nabla_{c}h^{cd}+\left(\nabla_{a}R_{db}\right)h^{da}\\
& & +R_{db}\nabla_{a}h^{da}+\left(\nabla_{a}R_{\,\,dcb}^{a}\right)h^{cd}+R_{\,\,dcb}^{a}\nabla_{a}h^{cd}\\
& = & \nabla_{b}\square h+\lambda m\nabla_{b}\nabla A+R_{bd}\nabla^{d}h+\lambda mR_{bd}A^{d}\\
& & +\left(\nabla_{a}R_{db}\right)h^{da}+R_{db}\nabla^{d}h+\lambda mR_{db}A^{d}\\
&&+\left(\nabla_{a}R_{\,\,dcb}^{a}\right)h^{cd}+R_{\,\,dcb}^{a}\nabla_{a}h^{cd}
\end{eqnarray*}
\item  
\begin{eqnarray*}
\nabla^{a}\nabla_{b}\nabla_{a}h  =  \nabla_{a}\nabla_{b}\nabla^{a}h= \nabla_{b}\square h+R_{ba}\nabla^{a}h
\end{eqnarray*}
\end{enumerate}
In the above calculations the symmetries of the Riemann tensor are used as well as the commutation relation between two covariant derivatives. 

For the spin-1 field, we can perform a very similar calculation which involves again the application of the covariant derivative

\begin{align}
\nabla^{b}{\mathcal{D}^{a}}_{b}A_{a} 
 = & \frac{1}{2}\nabla^{b}\left(R_{b}^{a}A_{a}\right)+\frac{1}{2}\lambda m^{2}\nabla A\nonumber \\
&-\frac{1}{2\zeta m^{2}}\nabla_{a}\left[R_{c}^{a}\nabla^{c}\nabla^{c}\left(R_{bc}A^{b}\right)\right]\nonumber\\
 & -\frac{\lambda}{2\zeta}\nabla^{b}\left[R_{bc}\nabla^{c}\nabla A\right]\nonumber \\
 & -\frac{1}{2\zeta}\nabla^{b}\left[\lambda\nabla_{b}\nabla^{c}\left(R_{ac}A^{a}\right)\right]\nonumber \\
 & -\frac{\lambda^{2}m^{2}}{2\zeta}\nabla^{b}\left[\nabla_{b}\nabla A\right]\label{eq:51}
\end{align}
By using the gauge condition from above given by $\nabla^{c}\left(A^{a}R_{ac}\right)=-\lambda m^{2}\nabla^a A_a-\zeta m^3\varphi$ we can transform the whole expression into the following form
\begin{align}
\nabla^{b}{\mathcal{D}^{a}}_{b}A_{a}  = & -\frac{1}{2}\left(\lambda m^{2}\nabla A+\zeta m^3\varphi\right)+\frac{1}{2}\lambda m^{2}\nabla A\nonumber \\
 & +\frac{1}{2\zeta m^{2}}\nabla^{b}\left[R_{bc}\nabla^{c}\left(\lambda m^{2}\nabla A+\zeta m^3\varphi\right)\right]\nonumber \\
 & -\frac{\lambda}{2\zeta}\nabla^{b}\left[R_{bc}\nabla^{c}\nabla A\right]\nonumber \\
 & +\frac{1}{2\zeta}\nabla^{b}\left[\lambda\nabla_{b}\left(\lambda m^{2}\nabla A+\zeta m^3\varphi\right)\right]\nonumber \\
 & -\frac{\lambda^{2}m^{2}}{2\zeta}\nabla^{b}\left[\nabla_{b}\nabla A\right]\nonumber \\
 = & -\frac{1}{4}\zeta m^3\varphi+\frac{1}{2m}\nabla^{b}\left[R_{bc}\nabla^{c}\varphi\right]+\frac{1}{2}\lambda m\Box\varphi\label{eq:52}
\end{align}
which yields the desired result for the Stückelberg scalar.


\bibliography{literaturr}

@article{yao88,
  author = {{York-Peng} Yao and C. Yuan},
  journal = {Physical Review D},
  number = 7,
  pages = {2237--2244},
  title = {Modification of the equivalence theorem due to loop corrections},
  volume = 38,
  year = 1988
 }

@article{buchbinder2000equations,
  title={Equations of motion for massive spin 2 field coupled to gravity},
  author={Buchbinder, Ioseph Lvovich and Gitman, Dmitri Maximovitch and Krykhtin, VA and Pershin, VD},
  journal={Nuclear Physics B},
  volume={584},
  number={1-2},
  pages={615--640},
  year={2000},
  publisher={Elsevier}
}

@article{ash75,
     title = {Quantum Fields in Curved Space-Times},
     author = {Ashtekar, A. and Magnon, Anne},
     journal = {Proceedings of the Royal Society of London. Series A, Mathematical and Physical Sciences},
     volume = {346},
     number = {1646},
     jstor_formatteddate = {Nov. 4, 1975},
     pages = {pp. 375-394},
      year = {1975},
}

@article{mandal2024ward,
  title={Ward identities under the frame transformations in curved space-time},
  author={Mandal, Susobhan},
  journal={Communications in Theoretical Physics},
  volume={76},
  number={9},
  pages={095406},
  year={2024},
  publisher={IOP Publishing}
}

@article{de2025gravitational,
  title={Gravitational memory and Ward identities in the local detector frame},
  author={De Luca, Valerio and Khoury, Justin and Wong, Sam SC},
  journal={Physical Review D},
  volume={112},
  number={2},
  pages={024032},
  year={2025},
  publisher={APS}
}

@article{friedman1992unitarity,
  title={Unitarity of interacting fields in curved spacetime},
  author={Friedman, John L and Papastamatiou, Nicolas J and Simon, Jonathan Z},
  journal={Physical Review D},
  volume={46},
  number={10},
  pages={4442},
  year={1992},
  publisher={APS}
}

@article{green2024goldstone,
  title={A Goldstone boson equivalence for inflation},
  author={Green, Daniel and Gupta, Kshitij and Huang, Yiwen},
  journal={Journal of High Energy Physics},
  volume={2024},
  number={9},
  pages={1--32},
  year={2024},
  publisher={Springer}
}

@article{fried92,
  title={Unitarity of interacting fields in curved spacetime},
  author={Friedman, John L and Papastamatiou, Nicolas J and Simon, Jonathan Z},
  journal={Physical Review D},
  volume={46},
  number={10},
  pages={4442},
  year={1992},
  publisher={APS}
}

@article{dmor09,
  title={Distinguished quantum states in a class of cosmological spacetimes and their Hadamard property},
  author={Dappiaggi, Claudio and Moretti, Valter and Pinamonti, Nicola},
  journal={Journal of mathematical physics},
  volume={50},
  number={6},
  pages={062304},
  year={2009},
  publisher={American Institute of Physics}
}

@article{ag15,
  title={Preferred instantaneous vacuum for linear scalar fields in cosmological space-times},
  author={Agullo, Ivan and Nelson, William and Ashtekar, Abhay},
  journal={Physical Review D},
  volume={91},
  number={6},
  pages={064051},
  year={2015},
  publisher={APS}
}

@article{dap09,
  title={Cosmological horizons and reconstruction of quantum field theories},
  author={Dappiaggi, Claudio and Moretti, Valter and Pinamonti, Nicola},
  journal={Communications in mathematical physics},
  volume={285},
  number={3},
  pages={1129},
  year={2009},
  publisher={Springer}
}

@article{hass12,
  title={Resolving the ghost problem in nonlinear massive gravity},
  author={Hassan, Sayed Fawad and Rosen, Rachel A},
  journal={Physical review letters},
  volume={108},
  number={4},
  pages={041101},
  year={2012},
  publisher={APS}
}

@article{bd72,
  title={Can gravitation have a finite range?},
  author={Boulware, David G and Deser, Stanley},
  journal={Physical Review D},
  volume={6},
  number={12},
  pages={3368},
  year={1972},
  publisher={APS}
}

@book{gsw1,
  title={Superstring Theory: Volume 1, Introduction},
  author={Green, Michael B and Schwarz, John H and Witten, Edward},
  year={2012},
  publisher={Cambridge University Press}
}

@article{he92,
  author = {{Hong-Jian} He and {Yu-Ping} Kuang and {Xiaoyuan} Li},
  journal = {Physical Review Letters},
  number = 18,
  pages = {2619--2622},
  title = {On the precise formulation of the equivalence theorem},
  volume = 69,
  year = 1992
 }

@article{he94,
  author = {{Hong-Jian} He and {Yu-Ping} Kuang and {Xiaoyuan} Li},
  journal = {Physical Review D},
  number = 9,
  pages = {4842-4872},
  title = {Further investigation on the precise formulation of the equivalence theorem},
  volume = 49,
  year = 1994
 }

@article{he97,
  author = {{Hong-Jian} He and {William B} Kilgore},
  journal = {Physical Review D},
  number = 3,
  pages = {1515-1532},
  title = {Equivalence theorem and its radiative-correction-free formulation for all $R_\xi$ gauges},
  volume = 55,
  year = 1997
 }

@article{bag90,
  author = {{Jonathan} Bagger and {Carl} Schmidt},
  journal = {Physical Review D},
  number = 1,
  pages = {264--270},
  title = {Equivalence theorem redux},
  volume = 41,
  year = 1990
 }

@article{ark03,
  author = {{Nima} Arkani--Hamed and {Howard} Georgi and {Matthew D.} Schwartz},
  journal = {Annals of Physics},
  number = 2,
  pages = {96--118},
  title = {Effective Field Theorie for Massive Gravitons and Gravity in Theory Space},
  volume = 305,
  year = 2003
 }

@article{hor97,
  author = {J. Ho\v{r}ej\v{s}\'i},
  journal = {Czechoslovak Journal of Physics},
  number = 10,
  pages = {951--1066},
  title = {Electroweak Interactions and High-Energy Limit},
  volume = 47,
  year = 1997
  }

@article{chan85,
  author = {Michael Chanowitz and Mary Gaillard},
  journal = {Nuclear Physics B},
  number = 10,
  pages = {951--1066},
  title = {The TeV Physics of Strongly Interacting W's and Z's},
  volume = 261,
  year = 1985
  }

@article{lee77,
  author = {{B.W.} Lee and {C.} Quigg and {H.} Thacker},
  journal = {Physical Review D},
  number = 5,
  pages = {1519--1531},
  title = {Weak Interactions at Very High Energies: The Role of the Higgs Boson Mass},
  volume = 16,
  year = 1977
  }

@article{higuchi1987forbidden,
  title={Forbidden mass range for spin-2 field theory in de Sitter spacetime},
  author={Higuchi, Atsushi},
  journal={Nuclear Physics B},
  volume={282},
  pages={397--436},
  year={1987},
  publisher={Elsevier}
}

@article{anderson1987adiabatic,
  title={Adiabatic regularization in closed Robertson-Walker universes},
  author={Anderson, Paul R and Parker, Leonard},
  journal={Physical Review D},
  volume={36},
  number={10},
  pages={2963},
  year={1987},
  publisher={APS}
}

@article{cuomo2020goldstone,
  title={Goldstone equivalence and high energy electroweak physics},
  author={Cuomo, Gabriel and Vecchi, Luca and Wulzer, Andrea},
  journal={SciPost Physics},
  volume={8},
  number={5},
  pages={078},
  year={2020}
}

@article{farzinnia2015prospects,
  title={Prospects for discovering the Higgs-like pseudo-Nambu-Goldstone boson of the classical scale symmetry},
  author={Farzinnia, Arsham},
  journal={Physical Review D},
  volume={92},
  number={9},
  pages={095012},
  year={2015},
  publisher={APS}
}

@article{parker1974adiabatic,
  title={Adiabatic regularization of the energy-momentum tensor of a quantized field in homogeneous spaces},
  author={Parker, Leonard and Fulling, Stephen A},
  journal={Physical Review D},
  volume={9},
  number={2},
  pages={341},
  year={1974},
  publisher={APS}
}

@article{PhysRevD.111.025019,
  title = {Local diagnostic program for unitary evolution in general spacetimes},
  author = {Choi, Ka Hei and Hofmann, Stefan and Schneider, Marc},
  journal = {Phys. Rev. D},
  volume = {111},
  issue = {2},
  pages = {025019},
  numpages = {21},
  year = {2025},
  month = {Jan},
  publisher = {American Physical Society},
  doi = {10.1103/PhysRevD.111.025019},
  url = {https://link.aps.org/doi/10.1103/PhysRevD.111.025019}
}

@article{lehm57,
  author = {H. Lehmann and K. Symanzik and W. Zimmermann},
  journal = {Il Nuovo Cimento},
  number = 2,
  pages = {319--333},
  title = {On the Formulation of Quantised Field Theories - II},
  volume = 6,
  year = 1957
  }

@article{delbourgo1988massive,
  title={Massive yang-mills theory: Renormalizability versus unitarity},
  author={Delbourgo, Robert and Twisk, S and Thompson, George},
  journal={International Journal of Modern Physics A},
  volume={3},
  number={02},
  pages={435--449},
  year={1988},
  publisher={World Scientific}
}

@article{corn74,
  author = {John Cornwall and David Levin and George Tiktopoulos},
  journal = {Physical Review},
  number = 4,
  pages = {1145--1167},
  title = {Derivation of Gauge Invariance From High-Energy Unitarity Bounds on the S-Matrix},
  volume = 10,
  year = 1974
  }

@article{dew75,
  author = {Bryce DeWitt},
  journal = {Physics Reports},
  number = 6,
  pages = {295--357},
  title = {Quantum Field Theory in Curved Spacetime},
  volume = 19,
  year = 1975
  }

@article{gou86,
  author = {{G.J.} Gounaris and {R.} K\"ogerler and {H.} Neufeld},
  journal = {Physical Review D},
  number = 10,
  pages = {295--357},
  title = {Relationship Between Polarised Vector Bosons and Their Scalar Partners},
  volume = 34,
  year = 1986
  }

@article{bir80,
  author = {{N.D.} Birrell and {J.G.} Taylor},
  journal = {Journal of Mathematical Physics},
  number = 7,
  pages = {1740--1760},
  title = {Analysis of Interacting Quantum Field Theory in Curved Spacetimes},
  volume = 21,
  year = 1980
  }

@book{bir82,
   author = {{N. D.} Birrell and {P. C. W.} Davies},
   title = {Quantum Fields in Curved Space},
   publisher = {Cambridge University Press},
   year = 1982
   }

@book{itz80,
   author = {Claude Itzykson and Jean-Bernard Zuber},
   title = {Quantum Field Theory},
   publisher = {Dover},
   year = 1980
   }

@article{berk10,
  author = {Felix Berkhahn and Dennis D. Dietrich and Stefan Hofmann},
  journal = {Journal of Cosmology and Astroparticle Physics},
  number = 11,
  title = {Self-Protection of Massive Cosmological Gravitons},
 volume = 2010,
  year = 2010
  }

@article{berk12,
  author = {Felix Berkhahn and Dennis D. Dietrich and Stefan Hofmann and Florian K\"uhnel and Parvin Moyassari},
  journal = {Physical Review Letters},
  number = 108,
  title = {Island of Stability  for Consistent Deformations of Einstein's Gravity},
  year = 2012
  }

@article{becch76,
  author = {C. Becchi and A. Rouet and R. Stora},
  journal = {Annals of Physics},
  number = 2,
  pages = {287--321},
  title = {Renormalisation of Gauge Theories},
  volume = 98,
  year = 1976
  }

@article{paul39,
  author = {M. Fierz and W. Pauli},
  journal = {Proceedings of the Royal Society},
  number = A,
  pages = {211--232},
  title = {On Relativistic Wave Equations for Particles of Arbitrary Spin in an Electromagnetic Field},
  volume = 173,
  year = 1939
  }

@article{derham,
  title={Resummation of massive gravity},
  author={De Rham, Claudia and Gabadadze, Gregory and Tolley, Andrew J},
  journal={Physical Review Letters},
  volume={106},
  number={23},
  pages={231101},
  year={2011},
  publisher={APS}
}

@article{hint12,
  author =  {Kurt Hinterbichler},
  journal = {Reviews of Modern Physics},
  number = 2,
  title = {Theoretical aspects of massive gravity},
  volume = 84,
  year = 2012
 }

\end{document}